\documentclass[]{article}
\usepackage{amsmath}
\usepackage{amsfonts}
\usepackage{amssymb}
\usepackage{graphicx}
\usepackage{amsfonts}
\usepackage[english]{babel}
\usepackage[utf8]{inputenc}
\usepackage{times}
\usepackage[T1]{fontenc}
\usepackage{multirow}
\usepackage{hyperref}

\begin{document}

\title{Hamiltonian and Diffeomorphism Constraints Generalized for Timelike and Spacelike 3+1 Foliation}

\author{Leonid Perlov, \\
Department of Physics, University of Massachusetts,  Boston, USA\\
leonid.perlov@umb.edu
}

\maketitle

\begin{abstract}
The form of Hamiltonian and Diffeomorphism constraints in Sen-Ashtekar-Barbero-Immirzi variables is well known for the spacelike 3+1 ADM foliation. It is also known that Sen-Ashtekar-Barbero-Immirzi connection can be introduced only in 3 dimensional space and does not work for $D > 3$. The reason it works in $D = 3$ is due to existence of isomorphism between $so(3)$ algebra and  $R^3$ space with a vector product. It turns out that similar isomorphism exists between $so(2,1)$ algebra and $R^3_{2,1}$ space algebra with respect to its vector product. By using this isomorphism we find both analog of Sen-Ashtekar-Barbero-Immirzi connection for timelike 3+1 foliation and the corresponding forms of Gauss, Diffeomorphism and Hamiltonian constraints. We then combine spacelike and timelike foliation constraints into the generalized form of the Hamiltonian and Diffeomorphism constrains using generalized Sen-Ashtekar-Barbero-Immirzi connection variables. We prove that Immirzi parameter is covariant with respect to timelike-spacelike ADM foliation change as in both cases in self-dual Ashtekar case it disappears in Hamiltionian constraint keeping it polynomial. 
\end{abstract}
 
\section{Introduction}
 It is known that Sen-Ashtekar-Barbero-Immirzi connection \cite{Barbero}, \cite{Immirzi} and flux variables can be introduced only in 3 dimensional space and do not work for $D > 3$, see $\cite{Thiemann}$. The reason it works in $D = 3$ is due to isomorphism between $so(3)$ algebra and  $R^3$ space with a vector product. Such isomorphism does not exist for $D > 3$, and, therefore it is impossible to introduce Sen-Ashtekar-Barbero-Immirzi connection.\\
We have noticed that similar isomorphism also exists in $D = 3$ between $so(2,1)$ algebra and algebra of vectors in $R^3_{2,1}$ space with its vector product \cite{Fomenko}. By using this isomorphism we derive Sen-Ashtekar-Barbero-Immirzi formalism in timelike foliation ADM with $SO(2,1)$ structure group. We obtain a new connection in that case and corresponding rotational, Gauss, Diffeomorphism and Hamiltonian constraints. We then combine $so(3)$ connection for spacelike foliation ADM with $so(2,1)$ conneciton for timelike foliation ADM into one expression, which we call a generalized connection. We continue by combining rotational, Gauss, Diffeomorphism and Hamiltonian set of constraints into one set of generalized constraints for both cases. \\
Recently 3+1 timelike foliation obtained much attention due to an attempt to make the next step towards covariant theory. A timelike foliation was considered in $\cite{Alexandrov-Kadar}$, $\cite{Alexandrov2}$, $\cite{Alexandrov3}$, $\cite{Montesinos}$, $\cite{Noui}$, providing new variables.\\
Our result is novel as we obtain the timelike case constraints in a much simpler form and directly from the original ADM variables by using isomorphism between $so(2,1)$ and $R^3_{2,1}$ algebras. Moreover, we were able to combine spacelike and timelike cases together into  generalized form by using the relation between $so(2,1)$ and $so(3)$ algebra structure coefficients.\\[2ex]
The paper is organized as follows. In section $\ref{sec:IsomorphismAndStructureCoefficients}$ we discuss  $so(2,1) \rightarrow R^3_{2,1}$ isomorphism.  Then in $\ref{sec:FoliationWithTimelikeSurfaces}$ we remind the formalism of the spacetime foliation by the timelike surfaces in original variables $P^{ab}, q_{ab}$. In the following section $\ref{sec:GeneralizedSO(2,1)-SO(3)RotationalConstraint}$ we introduce $SO(2,1)$ rotational constraints and extrinsic curvature variables $K^i_a$. We then write $SO(3)$ and $SO(2,1)$ rotational constraints in a combined generalized form. In section $\ref{sec:GeneralizedSen-AshtekarConnection}$ we obtain a generalized Sen-Ashtekar-Barbero-Immirzi connection for both timelike and spacelike ADM foliation as well as a generalized covariant derivative and a generalized Gauss constraint. In section $\ref{sec:SymplecticStructure}$ we introduce a generalized canonical transformation between variables $(q_{ab}, P^{cd})$ and Sen-Ashtekar-Barbero-Immirzi variables $(A^i_a, E^i_a)$. By calculating Poisson brackets we show that the symplectic structure is preserved, i.e the transformation is canonical. In section $\ref{sec:Diffeomorphism}$ we derive a diffeomorphism constraint for timelike ADM foliation. We then combine it with the spacelike ADM foliation into one generalized equation. In section $\ref{sec:Hamiltonian}$ we derive the generalized form of the Hamiltonian constraint. A number of appendices below show all calculation details. Everywhere below we use the following index convention for symmetrization and anti-symmetrization: $A_{(a} A_{b)} = \frac{1}{2}(A_a+A_b)$ , $A_{[a} A_{b]} = \frac{1}{2}(A_a-A_b)$

\section{$so(2,1) \rightarrow R^3_{2,1}$ Isomorphism and Structure coefficients}
\label{sec:IsomorphismAndStructureCoefficients}
Before considering $so(2,1) \rightarrow R^3_{2,1}$ isomorphism we will first remind how the similar isomorphism is written in spacelike ADM with $SO(3)$ structure group. The dynamics equations can be written in Sen-Ashtekar-Barbero-Immirzi polynomial form in that case only because of isomorphism between $so(3)$ algebra and algebra $R^3$ with vector product. This isomorphism makes $so(3)$ regular and adjoint representations the same, which is not the case for any dimension higher than 3, as such isomorphism exists only in $D =3$. The isomorphism can be written in the following form:
\begin{equation}
\label{SO(3)Isomorphism}
\Gamma^l_{ai} = {\epsilon_{\;ki}^l}_{so(3)}\Gamma^k_a
\end{equation}
,where ${\epsilon_{\;ki}^{l}}_{so(3)} = {\epsilon_{lki}}_{so(3)} = {\epsilon^{lki}}_{so(3)}$ - fully antisymmetric $so(3)$ tensor, which is also $so(3)$ algebra structure coefficients with $ {\epsilon^{012}}_{so(3)}=1,  {\epsilon_{012}}_{so(3)} = 1$, while $\Gamma^k_a$ are vectors in $R^3$. 
The covariant derivative can be written as:
\begin{equation}
\label{CDerSO3}
D_aE^a_i = [D_aE^a]_i + \Gamma^l_{ai}E^a_l = \partial_aE^a_i + {\epsilon^l_{\;ki}}_{so(3)}\Gamma^k_aE^a_l
\end{equation}
Similar isomorphism exists between $so(2,1)$ algebra and algebra of $R^3_{2,1}$ vectors with respect to its vector product:
\begin{equation}
\label{SO(2,1)Isomorphism}
\Gamma^l_{ai} = {\epsilon_{\;ki}^l}_{so(2,1)}\Gamma^k_a =  {\epsilon_{ik}^{\;\;\;l}}_{so(2,1)}\Gamma^k_a
\end{equation}
,where $ {\epsilon_{\;ki}^l}_{so(2,1)}$ is $so(2,1)$ antisymmetric tensor and algebra $so(2,1)$ structure coefficients with  $ {\epsilon^{012}}_{so(2,1)}=1,  {\epsilon_{012}}_{so(2,1)} = -1$, while $\Gamma^k_a$ are vectors in $R^3_{2,1}$.   \\[2ex]
Since $ {\epsilon_{\;ki}^l}_{so(2,1)} = \eta_{ij}{\epsilon_{k}^{jl}}_{so(3)}$, where $\eta_{ij} = Diag(-1,1,1)$, we can also rewrite it as:
\begin{equation}
\label{SO(2,1)Isomorphism}
\Gamma^l_{ai} = \eta_{ij}{\epsilon_k^{jl}}_{so(3)}\Gamma^k_a
\end{equation}
By using this isomorphism the covariant derivative can be written as:
\begin{multline}
\label{CDerSO(2,1)}
D_aE^a_i = [D_aE^a]_i + \Gamma^l_{ai}E^a_l = \partial_aE^a_i + {\epsilon^l_{ik}}_{so(2,1)}\Gamma^k_aE^a_l  =
   \partial_aE^a_i + \eta_{ij}{\epsilon_k^{jl}}_{so(3)}\Gamma^k_aE^a_l  
\end{multline}

\section{Space-time foliation with timelike surfaces }
\label{sec:FoliationWithTimelikeSurfaces}
The ADM foliation with spacelike surfaces is well known, see for example $\cite{Thiemann}, \cite{RovelliBook}$. The ADM foliation with timelike surfaces is less known, even though, it exists, see for example $\cite{Torii} , \cite{Alexandrov-Kadar}, \cite{Noui}$. In the last two papers it is obtained from the Holst action \cite{Holst}, while in the first one it's obtained in the original variables even for the more general case of Einstein-Gauss-Bonnet gravity. If one ingores the Gauss-Bonnet additional term, then it becomes the ADM timelike foliation of Einstein gravity. \\
We will repeate in brief this formalism in the original variables. The projected metric differs in spacelike and timelike foliations cases only by a sign s:
\begin{equation}
q_{\mu \nu} = g_{\mu \nu} - s n_{\mu} n_{\nu}
\end{equation}
,where $n_{\mu}$ is a unit normal vector to a foliation surface $\Sigma$  with $n_{\mu}n^{\mu} = s$. The vector $n_{\mu}$ is timelike and foliation is by spacelike surfaces when $s=-1$, and, correspondingly, $n_{\mu}$ is spacelike and the foliation is by timelike surfaces when $s = 1$.
Lagrangian expressed via projected metric is:\\
\begin{equation}
L = \; ^{(4)}\,R\sqrt{-\det(g)} = |q|^{1/2}N \; ^{(4)}R = |q|^{1/2}N( R - s( K_{ab}K^{ab} - K^2))
\end{equation}
,where N is the normal projection of the time coordinate in spacelike foliation and a normal projection of the space coordinate in the timelike foliation.
\begin{equation}
K_{ab} = \frac{1}{2N}({\dot{q}_{ab}} - (L_{\vec{N}}q)_{ab})
\end{equation}
ADM action for both timelike and spacelike case in $P^{ab}, q_{ab}$ variables is
\begin{equation}
S = \int d^4x \; |q|^{1/2}N( R - s( K_{ab}K^{ab} - K^2 ))
\end{equation}
\begin{equation}
P^{ab}= \frac{\partial L}{\partial {\dot{q}_{ab}}} = -s |q|^{1/2}(K^{ab} - tr(K) q^{ab})
\end{equation}
The Legendre transform then produces for both cases:
\begin{equation}
S = \int d^4x \; (\dot{q}_{ab}P^{ab} + N^aH_a + NH)
\end{equation}
, where
\begin{equation}
H_a = -2q_{ac}D_bP^{bc}
\end{equation}
\begin{equation}
\label{Hamiltonian22}
H = \frac{-s}{\sqrt{|q|}}\left[q_{ac}q_{bd} - \frac{1}{D-1}q_{ab}q_{cd}\right] P^{ab}P^{cd} - \sqrt{|q|} \;R
\end{equation}

\section{Generalized SO(2,1)-SO(3) Rotational Constraint}
\label{sec:GeneralizedSO(2,1)-SO(3)RotationalConstraint}
In this section we will obtain a new rotational constraint for timelike ADM foliation with $SO(2,1)$ structure group. 
We begin with $( q_{ab}, P^{cd})$ and  introduce the $SO(2,1)$ Sen-Ashtekar-Barbero-Immirzi variables in a canonical way. On four dimensional Lorentz manifold with 3+1 timelike foliation we introduce a bundle space with triads invariant with respect to $SO(2,1)$ rotation.
\begin{equation}
\label{q}
q_{ab} = e^i_a e^j_b \eta_{ij}
\end{equation}\\
,where $\eta_{ij}$ is Minkowski $R^3_{1,2}$ metric $\eta_{ij}= Diag(-1,1,1)$
\\[2ex]
We introduce an electric flux variable as a weight one density:
\begin{equation}
\label{E}
E^a_j = {|\det(e^i_a)|}e^a_j, \;\;  E^j_a = e^j_a/{|\det(e^i_a)|}
\end{equation}
We will use the notation $q = -|\det(e^i_a)|^2$\\[2ex]
We then introduce $K^i_a$ one-form in a little different way than in a spacelike case (notice $\eta_{ij}$):
\begin{equation}
\label{Kab}
K_{ab} := K^i_{(a}e^j_{b)}\eta_{ij}
\end{equation}
satisfying modified rotational constraint, again notice $\eta_{ij}$:
\begin{equation}
\label{rconstraint}
G_{ab} := K^i_{[a}e^j_{b]}\eta_{ij}  = 0
\end{equation}
By using ($\ref{E}$) we can rewrite it as:
\begin{equation}
\label{rconstraint1}
G_{ab} := K^i_{[a}E^j_{b]}\eta_{ij}  = 0
\end{equation}
or by raising indices a and b we obtain the form:
\begin{equation}
\label{rconstraint2}
G^{ab} := q^{at}q^{be}K^i_{[t}E^j_{e]}\eta_{ij}  = 0
\end{equation}
Notice that  ($\ref{q}$), ($\ref{E}$), ($\ref{Kab}$), ($\ref{rconstraint}$), ($\ref{rconstraint1}$) and ($\ref{rconstraint2}$)  differ from corresponding $SO(3)$ expressions by Minkowski metric $\eta_{ij}$ presence. \\[2ex]
The rotational constraint ($\ref{rconstraint1}$) can be converted into a different, although equivalent form, by contracting it with $e^a_k e^b_m$:
\begin{equation}
\label{jkconstraint}
G_{ab}e^a_k e^b_m = K^i_ae^j_b\eta_{ij}e^a_k e^b_m  - K^i_be^j_a\eta_{ij}e^a_k e^b_m  = K_{aj}e^a_k\delta^j_m - K_{bj}e^ b_m\delta^j_k = 2 K_{a[m}e^a_{k]} = 0
\end{equation}
,where we used $e^j_b e^b_m  = \delta^j_m$ and $K^i_a\eta_{ij} = K_{aj}$.  By using ($\ref{E}$) and changing index $m$ to $j$, it can also be written as:
\begin{equation}
\label{jkconstraint2}
G_{jk} = K_{a[j}E^a_{k]}=0
\end{equation}
or by using $SO(3)$ antisymmetric tensor as:
\begin{equation}
\label{RotConstraint1}
G^i = {\epsilon^{ijk}}_{so(3)} K_{aj}E^a_{k}=0
\end{equation}
Finally we can rewrite these constraints once again by contracting each $G^i$ with  $\eta_{ij}$.
\begin{equation}
\label{RotConstraint2}
G_i = \eta_{ij}{\epsilon^{jkl}}_{so(3)}K_{ak}E^a_{l}= {\epsilon_i^{\;\;kl}}_{so(2,1)}K_{ak}E^a_{l}=  {\epsilon_{ik}^{\;\;\;l}}_{so(2,1)}K^k_{a}E^a_{l}= 0
\end{equation}
We define now $\bar{\eta}_{ij}$ to be $Diag(-1,1,1)$ in $SO(2,1)$ case, and $Diag(1,1,1)$ in $SO(3)$ case, in order to write $SO(3)$ rotational constraint ($\ref{RotConstraint1}$) and $SO(2,1)$ rotational constraint ($\ref{RotConstraint2}$) in a generalized form:
\begin{equation}
\label{RotConstraint8}
G_i = \bar{\eta}_{ij}{\epsilon^{jkl}}_{so(3)}K_{ak}E^a_{l}=\bar{\eta}_{ij}{\epsilon_k^{jl}}_{so(3)}K^k_{a}E^a_{l}=0
\end{equation}

\section{Generalized SO(2,1)-SO(3) Sen-Ashtekar-Barbero-Immirzi Connection and Generalized Gauss Constraint }
\label{sec:GeneralizedSen-AshtekarConnection}
By using ($\ref{CDerSO3}$) and ($\ref{CDerSO(2,1)}$) and a generalized metric $\bar{\eta}_{ij}$ defined at the end of the previous section, we can write the generalized covariant derivative as:
\begin{equation}
\label{CDerGeneral}
D_aE^a_i = [D_aE^a]_i + \Gamma^l_{ai}E^a_l = \partial_aE^a_i + \bar{\eta}_{ij}{\epsilon_k^{jl}}_{so(3)}\Gamma^k_aE^a_l 
\end{equation}
,where $\Gamma^k_a$ are vectors correspondingly in $R^3_{2,1}$ for $SO(2,1)$ timelike foliation, and in $R^3$ for $SO(3)$ spacelike foliation. \\[2ex]
One can easily see that $R^3_{2,1}$ vectors $\Gamma^i_a$ are invariant under Weyl canonical transformation in $SO(2,1)$ timelike case similar to $SO(3)$ spacelike case:
\begin{equation}
(K^i_a, E^a_i) \rightarrow ^{(\beta)}K^i_a = \beta K^i_a, \;\; ^{(\beta)}E^a_i = E^a_i/\beta
\end{equation}
,where $\beta \in C$ is Barbero-Immirzi parameter.\\[2ex]
An invariance follows from the explicit formula for  $\Gamma^i_a$ expressed via $\Gamma^l_{aj}$ in ($\ref{SO(2,1)Isomorphism}$), when the latter is expressed via triads. Thus in both $SO(3)$ and $SO(2,1)$ case we can write it in a generalized form:
\begin{multline}
\Gamma^i_a = \frac{1}{2} \bar{\eta}^{im}{\epsilon_m^{jk}}_{so(3)}e^b_k[e^j_{a,b} - e^j_{b,a} + e^c_je^l_ae^l_{c,b}] 
\\
= \frac{1}{2}\bar{\eta}^{im}{\epsilon_m^{jk}}_{so(3)}E^b_k[E^j_{a,b} - E^j_{b,a} + E^c_jE^l_aE^l_{c,b}]
\\
=\frac{1}{4}\bar{\eta}^{im}{\epsilon_m^{jk}}_{so(3)}E^b_k\left[2E^j_a\frac{(\det(E))_{,b}}{\det(E)} - E^j_b \frac{(\det(E))_{,a}}{\det(E)}\right]
\end{multline}
We can see that $\Gamma^i_a$ is a homogeneous function of degree zero, therefore $^{(\beta)}\Gamma^j_a= \Gamma^j_a$ and a covariant derivative $D_a$ does not depend on $\beta$ and  $D_a(^{(\beta)}E^a_j) = 0$ in both $SO(3)$ \cite{Thiemann} and $SO(2,1)$ cases.  It is easy to see that the generalized rotational constraint ($\ref{RotConstraint8}$) also does not depend on $\beta$.\\[2ex]
Therefore, by using ($\ref{RotConstraint8}$) and the generalized covariant derivative ($\ref{CDerGeneral}$), we can write:
\begin{multline}
\label{result}
G_i = 0 + {\bar{\eta}}_{ij}{\epsilon_k^{jl}}_{so(3)}  (^{\beta} K^k_{a})  (^{\beta} E^a_l) 
\\
=D_a(^{(\beta)}E^a_i)  + {\bar{\eta}}_{ij}{\epsilon_k^{jl}}_{so(3)}  (^{\beta} K^k_{a})  (^{\beta} E^a_l) 
\\
=  \partial_aE^a_i + \bar{\eta}_{ij}{\epsilon_k^{jl}}_{so(3)}\Gamma^k_a (^{\beta} E^a_l) +  {\bar{\eta}}_{ij}{\epsilon_k^{jl}}_{so(3)}  (^{\beta} K^k_{a})  (^{\beta} E^a_l) 
\\
= \partial_aE^a_i + \bar{\eta}_{ij}{\epsilon_k^{jl}}_{so(3)}\Gamma^k_a (^{\beta} E^a_l) +  {\bar{\eta}}_{ij}{\epsilon_k^{jl}}_{so(3)}  (^{\beta} K^k_a)  (^{\beta} E^a_l) 
\\
= \partial_a (^{\beta}E^a_i ) +  \bar{\eta}_{ij}{\epsilon_k^{jl}}_{so(3)} \left[\Gamma^k_a + (^{\beta}K^k_a) \right] (^{\beta}E^a_l) = ^{\beta}\mathcal{D}_a (^{\beta}E^a_i) = 0
\end{multline}
or by introducing notation:
\begin{equation}
\label{so(2,1)connection}
^{\beta}A^l_{ai} = \bar{\eta}_{ij}{\epsilon_k^{jl}}_{so(3)} (\Gamma^k_a + (^{\beta}K^k_a))
\end{equation}
and notations:
\begin{equation}
\label{GeneralizedConnection}
^{\beta}A^k_a = \Gamma^k_a + (^{\beta}K^k_a),   \;\;\; ^{\beta}A^l_{ai} = \bar{\eta}_{ij}{\epsilon_k^{jl}}_{so(3)} A^k_a
\end{equation}
we can rewrite ($\ref{result}$) as a generalized Gauss constraint:
\begin{multline}
 ^{\beta}\mathcal{D}_a (^{\beta}E^a_i) =  \partial_a (^{\beta}E^a_i ) +(^{\beta}A^l_{ai})(^{\beta}E^a_l)  = \\
 \partial_a (^{\beta}E^a_i ) +  \bar{\eta}_{ij}{\epsilon_k^{jl}}_{so(3)} \left[\Gamma^k_a + (^{\beta}K^k_a) \right] (^{\beta}E^a_l)  = 0
\end{multline}

\section{Generalized Metric and Momentum Transformation to Sen-Ashtekar-Barbero-Immirzi Variables. Symplectic  Structure }
\label{sec:SymplecticStructure}
Before we go over to generalized $SO(2,1)-SO(3)$  Diffeomorphism and Hamiltonian constraints we have to define a new transformation between $q_{ab}, P^{cd}$ and Sen-Ashtekar-Barbero-Immirzi variables $E^a_i, K^j_a$ for timelike ADM foliation with $SO(2,1)$ structure group, and prove that symplectic structure is preserved, i.e in new variables we have the same dynamics. Such transformation is well known for $SO(3)$ case $\cite{Thiemann}$:
\begin{equation}
\label{pandqSO(3)}
q_{ab} = E^i_aE^j_b |\det E^c_i|^{2/D-1}, \;
P^{cd} = |\det(E^c_e)|^{-2/(D-1)} ( E^c_kE^{t}_{k}K^i_tE^d_i  - E^c_kE^{d}_{k}K^i_tE^t_i)
\end{equation}
Preserving symplectic structure
\begin{equation}
\label{Poissontrivial1}
\{E^a_j(x), E^b_k(y)\} = \{K^j_a(x), K^k_b(y)\} = 0, \; \{E^a_i(x), K^j_b(y)\} = \frac{k}{2}\delta^a_b\delta^j_i \delta(x,y)
\end{equation}
,where $k= 16\pi G/c^3$ - gravitational coupling constant.\\[2ex]
In order to define similar transformation for timelike $SO(2,1)$ foliation we need to modify ($\ref{pandqSO(3)}$) in the following way:
\begin{equation}
\label{pandqSO(2,1)}
q_{ab} = E^i_aE^j_b\eta_{ij} (-|\det E^c_i|^{2/D-1}), \;
P^{cd} = (-|\det(E^c_e)|^{-2/(D-1)}) ( E^c_kE^{t}_{m}\eta^{mk}K^i_tE^d_i  - E^c_kE^{d}_{m}\eta^{mk}K^i_tE^t_i)
\end{equation}
Notice  $\eta_{ij}$ presence in several places. \\[2ex]
We have to prove that this transformation preserves the same symplectic strucutre ($\ref{Poissontrivial1}$). Coordinate-coordinate Poisson bracket $\{q_{ab}, q_{cd}\}$ is zero, since $q_{ab}$ contains only electric fluxes $E^a_i$ and $\{E^a_j(x), E^b_k(y)\}$ is zero, as it follows from  ($\ref{Poissontrivial1}$) and ($\ref{pandqSO(2,1)}$). 
We calculate momentum-momentum Poisson bracket by using momentum formula ($\ref{pandqSO(2,1)}$):
\begin{equation}
P^{ab}(x) = (-|\det(E^c_e)|^{-2/(D-1)})( E^a_kE^{t}_{m}\eta^{mk}K^i_tE^b_i  - E^a_kE^{b}_{m}\eta^{mk}K^i_tE^t_i ) 
\end{equation}
\begin{equation}
P^{cd}(x) = (-|\det(E^c_e)|^{-2/(D-1)}) ( E^c_kE^{t}_{m}\eta^{mk}K^i_tE^d_i  - E^c_kE^{d}_{m}\eta^{mk}K^i_tE^t_i)
\end{equation}
\\
and notation $ q:= -|\det(E^c_e)|^{2/(D-1)}$ as follows:
\begin{multline}
\label{momentumbracket4}
\{P^{ab}(x), P^{cd}(y)\} = \{\frac{1}{q}(E^a_{k_1}E^{t_1}_{p_1}\eta^{p_1k_1}K^{i_1}_{t_1}E^b_{i_1} - E^a_{k_2}E^{b}_{p_2}\eta^{p_2k_2}K^{i_2}_{t_2}E^{t_2}_{i_2}) , \\
 \frac{1}{q}(E^c_{m_1}E^{e_1}_{p_3}\eta^{p_3m_1}K^{j_1}_{e_1}E^d_{j_1} - E^c_{m_2}E^{d}_{p_4}\eta^{p_4m_2}K^{j_2}_{e_2}E^{e_2}_{j_2})
\end{multline}
After lengthy calculations (see Appendix C) we obtain:
\begin{equation}
\label{momentumbracket12}
 \{P^{ab}(x), P^{cd}(y)\} = 2kq^{ac}G^{bd}
\end{equation}
,where \\[2ex]
$G^{bd} =q^{bt}q^{dp}K^{j}_{t}E^{i}_{p}\eta_{ij} - q^{dt}q^{bp}K^{j}_{t}E^{i}_{p}\eta_{ij}= K^{bj}E^{di}\eta_{ij} - K^{dj}E^{bi}\eta_{ij} =2K^{j[b}E^{d]i}\eta_{ij} = 2e^{bt}e^{de}K^{j}_{[t}E_{e]}^{i}\eta_{ij}  $\\[2ex]
is $SO(2,1)$ rotational constraint ($\ref{rconstraint2}$)  .\\[2ex]
When rotational constrain is zero, the Poisson bracket ($\ref{momentumbracket12}$) is also zero. Therefore momentum-momentum Poisson bracket remains on shell the same as in original variables $(q_{ab}, P^{cd})$. Exactly as in spacelike case.\\[2ex]
Finally we consider coordinate-momentum bracket:
\begin{equation}
P^{ab}(x) = (-|\det(E^c_e)|^{-2/(D-1)})( E^a_{k}E^{t}_{m}\eta^{mk}K^{i}_{t}E^b_{i} - E^a_{k}E^{b}_{m}\eta^{mk}K^{i}_{t}E^{t}_{i} ) 
\end{equation}
 \begin{equation}
 q_{cd}(y) = E^{j}_cE^{m}_d \eta_{mj}(-|\det(E^c_e)|)^{2/(D-1)}
 \end{equation}
 \begin{multline}
 \label{FirstPoissonBook}
\{P^{ab}(x),  q_{cd}(y)\} = |\det(E^c_e)|^{-2/(D-1)} \{( E^a_{k}E^{t}_{m}\eta^{mk}K^{i}_{t}E^b_{i} - E^a_{k}E^{b}_{m}\eta^{mk}K^{i}_{t}E^{t}_{i} ), \; E^j_cE^m_d \eta_{mj} |\det(E^c_e)|^{2/(D-1)}\} 
 \end{multline}
 After lengthy calculations (see Appendix C) we obtain: 
 \begin{equation}
  \{P^{ab}(x),  q_{cd}(y)\} =  k\delta^b_{(c}\delta^a_{d)}\delta(x, y)
 \end{equation}
so, the symplectic structure is preserved, i.e. new variables $E^a_{i}$ and $K^{i}_{a}$ are canonical in $SO(2,1)$ timelike foliation case as well. \\[2ex]
What remains is to write generalized transformations in both $SO(3)$ and $SO(2,1)$ cases. By  looking at ($\ref{pandqSO(2,1)}$) we see that it turns into ($\ref{pandqSO(3)}$), when instead of Minkowski metric $\eta_{ij}= Diag(-1,1,1)$ we use Euclidean $Diag(1,1,1)$. Therefore, by using the generalized metric $\bar\eta_{ij}$ we write generalized transformations:
\begin{equation}
\label{Generalized}
q_{ab} = E^i_aE^j_b\bar{\eta}_{ij}(- |\det E^c_i|^{2/D-1}), \;
P^{cd} =(-|\det(E^c_e)|^{-2/(D-1)}) ( E^c_kE^{t}_{m}\bar{\eta}^{mk}K^i_tE^d_i  - E^c_kE^{d}_{m}\bar{\eta}^{mk}K^i_tE^t_i)
\end{equation}
The symplectic structure ($\ref{Poissontrivial1}$) is the same in both cases so it can also be called generalized.

\section{Generalized SO(2,1)-SO(3) Diffeomorphism Constraint}
\label{sec:Diffeomorphism}
The diffeomorphism constraint in the original ADM variables can be written as in $\cite{Thiemann}$ (1.2.6):
\begin{equation}
\label{diffeomorphism}
H_a = -2sq_{ac}D_bP^{bc}
\end{equation}
where $s = -1$ for $SO(3)$ spacelike foliation case, while $s = 1$ in $SO(2,1)$ timelike case.\\[2ex]
By substituting generalized variables ($\ref{Generalized}$) into ($\ref{diffeomorphism}$) we obtain:
\begin{equation}
H_a = -2sq_{ac}D_b (-|\det(E^c_e)|^{-2/(D-1)} ( E^b_kE^{t}_{m}\bar{\eta}^{mk}K^i_tE^c_i  - E^b_kE^{c}_{m}\bar{\eta}^{mk}K^i_tE^t_i))
\end{equation}
or rewriting it by using metric expression: $q^{bt} =  E^b_kE^{t}_{m}\bar{\eta}^{mk}$ as 
\begin{equation}
H_a = -(2s/q)D_b ( q_{ac} q^{bt} q K^i_tE^c_i  - q_{ac}q^{bc} q K^i_tE^t_i) = -2sD_b ( K^{bi}E_{ai} - \delta^b_a  K^i_tE^t_i)
\end{equation}
we obtain a diffeomorphism constraint in variables $K^i_t$ and $E^t_i$:
\begin{equation}
\label{Diffeomorphism1}
 H_a = -2sD_b ( K^{bi}E_{ia} - \delta^b_a  K^i_tE^t_i) = -2sD_b ( K^i_{b}E^a_{i} - \delta^b_a  K^i_tE^t_i) 
\end{equation}
,where we at the same time lowered b and raised a in the first term. \\
By using an equality that follows from the rotational constraint ($\ref{rconstraint1}$):
we can show that (see Appendix G):
\begin{equation}
 K^i_{b}E^a_{i} =  K^{i}_aE^b_{i} 
\end{equation}
by substituting it into the first term of ($\ref{Diffeomorphism1}$) we rewrite it as:
\begin{equation}
\label{Diffeomorphism2}
 H_a =  -2sD_b ( K^i_{a}E^b_{i} - \delta^b_a  K^i_tE^t_i) 
\end{equation}
In order to express this constraint via generalized connection $A^i_a$ and electric flux $E^a_i$ we introduce generalized curvatures:
\begin{equation}
\label{Riemann1}
R^i_{ab} = 2\partial_{[a}\Gamma^i_{b]} + \bar{\eta}^{ij}{\epsilon_{jkl}}_{so(3)}\Gamma^{k}_a\Gamma^l_b
\end{equation}
and
\begin{equation}
\label{Riemann22}
^{(\beta)}F^i_{ab} = 2\partial_{[a} ^{(\beta)}A^i_{b]} +  \bar{\eta}^{ij}{\epsilon_{jkl}}_{so(3)} \; ^{(\beta)}A^k_a  \; ^{(\beta)}  A^l_b
\end{equation}
,where $A^k_a =\Gamma^k_a + \beta K^k_{a}$ is $so(3)$ connection in spacelike foliation, and $so(2,1)$ connection in timelike foliation. See ($\ref{Riemann1}$) and ($\ref{Riemann22}$) derivation in Appendix D.  \\[2ex]
By expressing $^{(\beta)}F^j_{ab} $ via $R^j_{ab} $, (see derivation in Appendix E) we obtain:
\begin{equation}
\label{ExtrinsicCurvature}
^{(\beta)}F^i_{ab} = R^i_{ab} + 2\beta D_{[a} K^i_{b]} + \beta^2 \bar{\eta}^{ij}{\epsilon_{jkl}}_{so(3)} K^k_a K^l_b
\end{equation}
By contracting ($\ref{ExtrinsicCurvature}$) with $^{(\beta)} E^b_i= E^b_i / \beta $ (see all calculations in Appendix F) we obtain:
\begin{equation}
\label{Contracting}
^{(\beta)}F^i_{ab} \;  ^{(\beta)} E^b_i = \frac{R^i_{ab} E^b_i}{\beta}  + 2D_{[a}K^i_{b]}E^b_i + \beta K^i_aG_j
\end{equation}
As in $so(3)$ case the first term on the right hand side is zero (see Appendix K). The last term is also zero on a shell, where the rotational constraint is zero. As for the second term, we write it as:
\begin{multline}
\label{Diffeomorphism3}
2D_{[a}K^i_{b]}E^b_{i}= D_a(K^i_{b})E^b_{i} - D_b(K^i_{a})E^b_{i} = - ( D_b(K^i_{a})E^b_{i}  - D_a(K^i_{b})E^b_{i}) ) = \\
- ( D_b(K^i_{a})E^b_{i}  - D_a(K^i_{b})E^b_{i} ) = \\
 - ( D_b(K^i_{a})E^b_{i}  - \delta^a_b D_a(K^i_{b})E^b_{i}) = \\
 - ( D_b(K^i_{a})E^i_{b}  - \delta^a_b D_a(K^i_b)E^b_i)  = \\
  - ( D_b(K^i_{a}E^b_{i})  - \delta^a_b D_a(K^i_bE^b_i)) = \\
   - (D_b(K^i_{a}E^b_{i})  - \delta^a_b D_a(K^i_bE^b_i)) = (-s/2)H_a 
\end{multline}
,where in order to go from the forth to the fifth line we used Lemma1 from Appendix A, while in the last line we used ($\ref{Diffeomorphism2}$): $ H_a = -2sD_b ( K^i_{a}E^b_{i} - \delta^b_a  K^i_tE^t_i)$. \\[2ex]
If then follows from ($\ref{Contracting}$) and ($\ref{Diffeomorphism3}$) that diffeomorphism constraint has the following form:
\begin{equation}
\label{Contracting1}
^{(\beta)}F^i_{ab} \;  ^{(\beta)} E^b_i = - (s/2)H_a + \beta K^i_aG_j
\end{equation}
or on shell:
\begin{equation}
H_a = -2s^{(\beta)}F^j_{ab} \;  ^{(\beta)} E^b_j
\end{equation}

\section{Generalized SO(2,1)-SO(3) Hamiltonian Constraint}
\label{sec:Hamiltonian}
Let us derive a Hamiltonian constraint. We will remind first how it was derived in $SO(3)$ spacelike ADM foliation \cite{Thiemann} case, and then we will derive it for $SO(2,1)$ timelike ADM foliation case. \\[2ex]
\textbf{$SO(3)$ Spacelike Foliation Case:}\\[2ex]
By contracting ($\ref{ExtrinsicCurvature}$) with ${\epsilon_j^{\;\;kl}}_{so(3)} {^{(\beta)}E^{a}_{k}}{^{(\beta)}E^{b}_{l}}$ we obtain:
\begin{multline}
\label{ContractingHamiltonianSO(3)1}
^{(\beta)}F^j_{ab} \; {\epsilon_j^{\;\;kl}}_{so(3)} {^{(\beta)}E^{a}_{k}}{^{(\beta)}E^{b}_{l}}  = -q \frac{R^j_{ab}{\epsilon^{kl}_{j}}_{so(3)}e^a_k e^b_l}{\beta^2}  + 
2\beta D_{[a} K^j_{b]} {\epsilon_j^{\;\;kl}}_{so(3)} {^{(\beta)}E^{a}_{k}}{^{(\beta)}E^{b}_{l}}  +
\\ {\beta^2 }{\epsilon^j_{mn}}_{so(3)} K^m_a K^n_b {\epsilon_j^{\;\;kl}}_{so(3)} {^{(\beta)}E^{a}_{k}}{^{(\beta)}E^{b}_{l}} 
\end{multline}
,where minus in the first term on the right hand side is because we have moved index $j$ by one position into the middle. \\
By using $R_{ab}^{\;\;\;kl} = R^j_{ab}{\epsilon^{kl}_{j}}_{so(3)}$ (see derivation in Appendix D) we rewrite the first term. By using (Appendix I) we rewrite the second term, and by using (Appendix J) we rewrite the third term as:
\begin{equation}
\label{ContractingHamiltonianSO(3)}
^{(\beta)}F^j_{ab} \; {\epsilon_{j}^{\;\;kl}}_{so(3)} {^{(\beta)}E^{a}_{k}}{^{(\beta)}E^{b}_{l}}  = -q \frac{R_{ab}^{\;\;\;kl}e^a_k e^b_l}{\beta^2} - 2 \frac{E^a_jD_aG_j}{\beta} + (K^j_aE^a_j)^2 - (K^j_bE^a_j)(K^k_aE^b_k)
\end{equation}
or, when contracting Riemann tensor with triads $ R_{ab}^{\;\;\;kl}e^a_k e^b_l = R$ we obtain:
\begin{equation}
\label{ContractingHamiltonianSO(3)1}
^{(\beta)}F^j_{ab} \; {\epsilon_{j}^{\;\;kl}}_{so(3)} {^{(\beta)}E^{a}_{k}}{^{(\beta)}E^{b}_{l}}  = -q \frac{R}{\beta^2} - 2 ^{(\beta)}E^a_jD_aG_j + (K^j_aE^a_j)^2 - (K^j_bE^a_j)(K^k_aE^b_k)
\end{equation}
or
\begin{multline}
\label{ContractingHamiltonianSO(3)2}
^{(\beta)}F^j_{ab} \; {\epsilon_{j}^{\;\;kl}}_{so(3)} {^{(\beta)}E^{a}_{k}}{^{(\beta)}E^{b}_{l}}  + 2 ^{(\beta)}E^a_jD_aG_j \\
= \frac{\sqrt{q}}{\beta^2}\left[ -\sqrt{q}R - \beta^2 \frac{(K^j_bE^a_j)(K^k_aE^b_k) - (K^j_aE^a_j)^2 }{\sqrt{q}}\right]\\
=  \frac{\sqrt{q}}{\beta^2}\left[ H + (s - \beta^2) \frac{(K^j_bE^a_j)(K^k_aE^b_k) - (K^j_aE^a_j)^2 }{\sqrt{q}}\right]
\end{multline}
,where in order to go from the second to the third line, we substituted the expression for $-\sqrt{q}R$ from the expression for Hamiltonian in $SO(3)$ case (see $\cite{Thiemann}$ 4.2.7):
\begin{equation}
\label{hamiltonian}
H = - \frac{s}{\sqrt{q}} (K^l_aK^j_b - K^j_aK^l_b)E^a_jE^b_l - \sqrt{q}R 
\end{equation}
It can be easily done when regrouping terms as follows:
\begin{multline}
\label{hamiltonian24}
H = - \frac{s}{\sqrt{q}} (K^l_aE^b_l K^j_bE^a_j - K^j_aE^a_jK^l_bE^b_l) - \sqrt{q}R = - \frac{s}{\sqrt{q}} ((K^j_aE^b_j) (K^k_bE^a_k) - (K^j_aE^a_j)^2) - \sqrt{q}R
\end{multline}
thus we obtain:
\begin{equation}
- \sqrt{q}R = H + - \frac{s}{\sqrt{q}} ((K^j_aE^b_j) (K^k_bE^a_k) - (K^j_aE^a_j)^2)
\end{equation}
We now express $H$ in ($\ref{ContractingHamiltonianSO(3)2}$):
\begin{multline}
H = \frac{\beta^2}{\sqrt{q}}\left[^{(\beta)}F^j_{ab} \; {\epsilon_{j}^{\;\;kl}}_{so(3)} {^{(\beta)}E^{a}_{k}}{^{(\beta)}E^{b}_{l}}  + 2 ^{(\beta)}E^a_jD_aG_j\right] \\
+ (\beta^2 - s)  \frac{(K^j_bE^a_j)(K^k_aE^b_k) - (K^j_aE^a_j)^2 }{\sqrt{q}}
\end{multline}
or
\begin{equation}
\label{HamiltonianSO(3)}
H = [{\beta^2} ^{(\beta)}F^j_{ab} - (\beta^2 - s) {\epsilon_{jmn}}_{so(3)} K^m_a K^n_b] \frac{{\epsilon_{j}^{\;\;kl}}_{so(3)} E^a_k E^b_l }{\sqrt{q}}
\end{equation}\\[2ex]
We have repeated $\cite{Thiemann}$ $SO(3)$ spacelike ADM foliation case for instructional aim.
We will now derive a new Hamiltonian constraint for timelike ADM foliation with $SO(2,1)$ structure group:\\[2ex]
\textbf{$SO(2,1)$ Timelike Foliation Case:}\\[2ex]
By contracting  ($\ref{ExtrinsicCurvature}$) now with ${\epsilon_{j}^{\;\;kl}}_{so(2,1)} {^{(\beta)}E^a_{k}}{^{(\beta)}E^b_{l}}$, we obtain (see details in Appendicies D, I and J):
\begin{equation}
\label{ContractingHamiltonianSO(2,1)1}
^{(\beta)}F^j_{ab} \; {\epsilon_j^{\;\;kl}}_{so(2,1)} {^{(\beta)}E^a_{k}}{^{(\beta)}E^b_{l}} = -q \frac{R}{\beta^2} - 2 \frac{E^a_{j}D_aG_j}{\beta} - ( (K^j_aE^a_j)^2 - (K^j_bE^a_j)(K^k_aE^b_k))
\end{equation}
By comparing ($\ref{ContractingHamiltonianSO(2,1)1}$) with ($\ref{ContractingHamiltonianSO(3)1}$), we see only one, however crucial difference: a different sign in the third term. \\[2ex]
By using ($\ref{ContractingHamiltonianSO(2,1)1}$)
\begin{multline}
\label{ContractingHamiltonianSO(2,1)2}
^{(\beta)}F^j_{ab} \; {\epsilon_j^{\;\;kl}}_{so(2,1)} {^{(\beta)}E^a_{k}}{^{(\beta)}E^b_{l}}  + 2 ^{(\beta)}E^a_{j}D_aG_j \\
= \frac{\sqrt{|q|}}{\beta^2}\left[ -\sqrt{|q|}R + \beta^2 \frac{(K^j_bE^a_j)(K^k_aE^b_k) - (K^j_aE^a_j)^2 }{\sqrt{|q|}}\right]\\
=  \frac{\sqrt{|q|}}{\beta^2}\left[ H + ( s + \beta^2 ) \frac{(K^j_bE^a_j)(K^k_aE^b_k) - (K^j_aE^a_j)^2 }{\sqrt{|q|}}\right]
\end{multline}
,where in order to go from the second to the third line, we substituted the expression for $\sqrt{q}R$ from Hamiltonian expression in case of timelike foliation (see Appendix B for details):
\begin{equation}
\label{hamiltonian}
H = - \frac{s}{\sqrt{|q|}} (K^l_aK^j_b - K^j_aK^l_b)E^a_jE^b_l - \sqrt{|q|}R 
\end{equation}
Notice the only, however crucial change in ($\ref{ContractingHamiltonianSO(2,1)2}$) compared to ($\ref{ContractingHamiltonianSO(3)2}$). We have $(s+ \beta^2)$ instead of $(s -\beta^2)$. This is due to the sign difference in the third term of ($\ref{ContractingHamiltonianSO(2,1)1}$) and ($\ref{ContractingHamiltonianSO(3)1}$), which is in turn a consequence of the Riemann tensor contraction with ${\epsilon_{jkl}}_{so(2,1)}$ vs ${\epsilon^{jkl}}_{so(3)}$, since: 
\begin{equation}
{\epsilon_{jkl}}_{so(3)}{\epsilon^{jmn}}_{so(3)} = 2\delta^{[m}_k\delta^{n]}_l
\end{equation}
while
\begin{equation}
{\epsilon_{jkl}}_{so(2,1)}{\epsilon^{jmn}}_{so(2,1)} = - 2\delta^{[m}_k\delta^{n]}_l
\end{equation}
By expressing H in ($\ref{ContractingHamiltonianSO(2,1)2}$) we obtain:
\begin{multline}
H = \frac{\beta^2}{\sqrt{|q|}}\left[^{(\beta)}F^j_{ab} \; {\epsilon_{j}^{\;\;kl}}_{so(2,1)} {^{(\beta)}E^a_{k}}{^{(\beta)}E^b_{l}}  + 2 ^{(\beta)}E^a_{j}D_aG_j\right] \\
- (\beta^2 + s)  \frac{(K^j_bE^a_j)(K^k_aE^b_k) - (K^j_aE^a_j)^2 }{\sqrt{|q|}}
\end{multline}
or
\begin{equation}
\label{HamiltonianSO(2,1)}
H = [{\beta^2} ^{(\beta)}F^j_{ab} - (\beta^2 + s) {\epsilon^j_{mn}}_{so(2,1)} K^m_a K^n_b] \frac{{\epsilon_j^{\;\;kl}}_{so(2,1)} E^a_k E^b_l }{\sqrt{|q|}}
\end{equation}\\[2ex]
We can now rewrite Hamiltonian constraints in ($\ref{HamiltonianSO(3)}$) and ($\ref{HamiltonianSO(2,1)}$) for $SO(3)$ and $SO(2,1)$ foliations in a general form:
\begin{equation}
\label{Generalized1}
H = [{\beta^2} ^{(\beta)}F^j_{ab} - (\beta^2 + 1) {\bar{\eta}^{ji}{\epsilon_{imn}}}_{so(3)} K^m_a K^n_b] \frac{\bar{\eta}_{ji}{\epsilon^{ikl}}_{so(3)}E^a_k E^b_l }{\sqrt{|q|}} 
\end{equation}\\[2ex]
,where, by remembering that $s = 1$ in $SO(2,1)$ and $s = -1$ in $SO(3)$ case, we have combined both $(\beta^2-s), s = -1$ in ($\ref{HamiltonianSO(3)}$) and  $(\beta^2+s), s = 1$ in ($\ref{HamiltonianSO(2,1)}$) into one expression $(\beta^2+1)$. We also used the following identities: $ {\bar{\eta}^{ji}{\epsilon_{imn}}}_{so(3)} = {\epsilon^j_{mn}}_{so(2,1)}$ in $SO(2,1)$ case and $ {\bar{\eta}_{ji}{\epsilon^{ikl}}}_{so(3)} = {\epsilon_j^{\;\;kl}}_{so(3)}$ in $SO(3)$ case.\\[2ex]
We can see in ($\ref{Generalized1}$) that the self-dual Ashtekar case for Immirzi $\beta = \pm i$ is preserved in both $SO(3)$ and $SO(2,1)$ cases, making general formula very simple:
\begin{equation} 
H = [ F^j_{ab}\bar{\eta}_{ji}{\epsilon^{ikl}}_{so(3)}E^a_k E^b_l] = 0
\end{equation}\\
This fact is very important as it proves that Immirzi parameter is covariant with respect to spacelike - timelike foliation change.  \\[2ex]
Finally we need to check that the hamiltonian and diffeomorphism constraints in new variables commute with the smeared rotational constraint. Like in \cite{Thiemann} we introduce the smeared constraint by using $G_{ik}$ form ($\ref{jkconstraint2}$)
\begin{equation}
G(\Lambda) = \int_{\sigma} d^3x \Lambda^{jk}K_{aj}E^a_k 
\end{equation}
, where $\Lambda \in so(3) $ for spacelike foliation and $\in so(2,1)$ for timelike foliation. 
The constraint satisfy the Poisson algebra:
\begin{equation}
\{G(\Lambda), G(\Lambda')\} = \frac{k}{2} G([\Lambda, \Lambda'])
\end{equation}
Since coordinate and momentum in ($\ref{Generalized}$) are $so(3)$ invariant in $SO(3)$ spacelike foliation case and $so(2,1)$ invariant in $SO(2, 1)$ timelike foliation case, they will commute with the corresponding smeared rotational constraint. Also both diffeormorphism and hamiltonian constraints commute with the smeared rotational constraint, since they are both functions of $q_{ab}$ and $P^{cd}$. So the whole system of constraints is still first class. 

\section{ All Generalized Constraints }
\label{sec:AllGeneralized Constraints}
To summarize we write the system of the generalized $SO(3)-SO(2,1)$ constraints together:
\begin{multline}
 ^{\beta}\mathcal{D}_a (^{\beta}E^a_i) = \partial_a (^{\beta}E^a_i ) +  \bar{\eta}_{ij}{\epsilon_k^{jl}}_{so(3)} \left[\Gamma^k_a + (^{\beta}K^k_a) \right] (^{\beta}E^a_l) = 0 \\
 \\
 H_a = -2s^{(\beta)}F^i_{ab} \;  ^{(\beta)} E^b_i = 0\\\\
 H = [{\beta^2} ^{(\beta)}F^i_{ab} - (\beta^2 + 1) {\bar{\eta}^{ij}{\epsilon_{jmn}}}_{so(3)} K^m_a K^n_b] \frac{\bar{\eta}_{ij}{\epsilon^{jkl}}_{so(3)}E^a_k E^b_l }{\sqrt{|q|}}  = 0
\end{multline}

\section{ Discussion }
\label{sec:Discussion}
The existence of isomorphism between $so(2,1)$ algebra and algebra of vectors in $R^3_{2,1}$ space with vector product as algebra operation was noticed and used to derive  Sen-Ashtekar-Barbero-Immirzi formalism for timelike foliation with $SO(2,1)$ structure group. A new $so(2,1)$ connection along with the Gauss, Diffeomorphism and Hamiltonian constraints have been obtained in $SO(2,1)$ case. The constraints in $SO(3)$ spacelike and in $SO(2,1)$ timelike ADM were combined into one set of generalized constraints using the generalized connection. In addition, it's been proved that Immirzi parameter is covariant with respect to timelike-spacelike ADM foliation change as in both cases in self-dual Ashtekar case it disappears in Hamiltionian constraint keeping it polynomial. 
 \\[3ex]

\section{Appendix A Lemma 1 for Diffeomorphism Constraints}
\label{sec:AppA}
\textbf{Lemma 1}
\begin{equation}
(D_a(K_{jb}) - D_b(K_{ja}))E^b_j = D_a(K_{jb}E^b_j) - D_b(K_{ja}E^b_j)
\end{equation}
\textbf{Proof:}
\begin{multline}
D_a(K_{jb}E^b_j) - D_b(K_{ja}E^b_j) = \delta^b_a D_b(K_{jb})E^b_j + \delta^b_aK_{jb}D_b(E^b_j) - D_b(K_{ja}) E^b_j - K_{ja}D_b(E^b_j) = \\
\delta^b_aD_b(K_{jb})E^b_j - D_b(K_{ja})E^b_j = (D_a(K_{jb}) - D_b(K_{ja}))E^b_j
\end{multline}
above in the first identity on the right hand side the second and the forth terms cancel. 

\section{Appendix B Hamiltonian Constraint in SO(2,1) Case}
We would like to express Hamiltonian constraint first in $K^i_a$ and $E^a_i$ variables and then in $A^i_a$, $E^a_i$ variables. We begin with Hamiltonian constraint ADM expression ($\ref{Hamiltonian22}$):
\label{sec:AppendixB}
\begin{equation}
H = \frac{-s}{\sqrt{|q|}}\left[q_{ac}q_{bd} - \frac{1}{D-1}q_{ab}q_{cd}\right] P^{ab}P^{cd} - \sqrt{|q|}R
\end{equation}
by substituting into it the metric $q_{ab}$ and momentum $P^{ab}$ expressions from ($\ref{pandqSO(2,1)}$) we obtain:
\begin{multline}
H = \frac{-s}{\sqrt{|q|}}\left[q_{ac}q_{bd} - \frac{1}{D-1}q_{ab}q_{cd}\right] \frac{1}{q}\left( q q^{at_1}K^{i_1}_{t_1}E^b_{i_1} - q q^{ab}K^{i_2}_{t_2}E^{t_2}_{i_2}\right) \\
\frac{1}{q}\left(qq^{ct_3}K^{i_3}_{t_3}E^d_{i_3} - qq^{cd}K^{i_4}_{t_4}E^{t_4}_{i_4}\right)  - \sqrt{|q|}R= \\
 \frac{-s}{\sqrt{|q|}}\left[q_{ac}q_{bd} - \frac{1}{D-1}q_{ab}q_{cd}\right] \left( K^{ai_1}E^b_{i_1} -  q^{ab}K^{i_2}_{t_2}E^{t_2}_{i_2}\right) \\
\left(K^{ci_3}E^d_{i_3} - q^{cd}K^{i_4}_{t_4}E^{t_4}_{i_4}\right)  - \sqrt{|q|}R
\end{multline}
by opening parentheses:
\begin{multline}
H = \frac{-s}{\sqrt{|q|}}\left[q_{ac}q_{bd} - \frac{1}{D-1}q_{ab}q_{cd}\right](K^{ai_1}E^b_{i_1}K^{ci_3}E^d_{i_3} -   q^{cd}K^{ai_1}E^b_{i_1}K^{i_4}_{t_4}E^{t_4}_{i_4} \\
 - q^{ab}K^{i_2}_{t_2}E^{t_2}_{i_2}K^{ci_3}E^d_{i_3} +  q^{ab}q^{cd}K^{i_2}_{t_2}E^{t_2}_{i_2}K^{i_4}_{t_4}E^{t_4}_{i_4})  - \sqrt{|q|}R= \\
  \frac{-s}{\sqrt{|q|}}( ( K^{i_1}_c E_{di_1}K^{ci_3}E^d_{i_3} - q_{ab}K^{ai_1}E^b_{i_1}K^{i_4}_{t_4}E^{t_4}_{i_4} - q_{cd}K^{i_2}_{t_2}E^{t_2}_{i_2}K^{ci_3}E^d_{i_3} + \\
q_{cd}q^{cd}K^{i_2}_{t_2}E^{t_2}_{i_2}K^{i_4}_{t_4}E^{t_4}_{i_4}) - \frac{1}{D-1} (K^{i_1}_bE^b_{i_1}K^{i_3}_dE^d_{i_3} - D K^{i_1}_bE^b_{i_1}K^{i_4}_{t_4}E^{t_4}_{i_4} - DK^{i_2}_{t_2}E^{t_2}_{i_2}K^{i_3}_dE^d_{i_3} -D^2K^{i_2}_{t_2}E^{t_2}_{i_2}K^{i_4}_{t_4}E^{t_4}_{i_4} ))  \\
- \sqrt{q}R = \frac{-s}{\sqrt{|q|}} (( K^{i_1}_c K^{ci_3}E_{di_1}E^d_{i_3} - K^{i_1}_bE^b_{i_1}K^{i_4}_{t_4}E^{t_4}_{i_4} - K^{i_2}_{t_2}E^{t_2}_{i_2}K^{i_3}_dE^d_{i_3} + 
DK^{i_2}_{t_2}E^{t_2}_{i_2}K^{i_4}_{t_4}E^{t_4}_{i_4}) \\
 - \frac{(1-D)^2}{D-1}(K^{i_1}_bE^b_{i_1}K^{i_3}_dE^d_{i_3})) -\sqrt{|q|}R = \frac{-s}{\sqrt{|q|}} ( K^{i_1}_c K^{c}_{i_3}E_{d}^{i_1}E^d_{i_3} + (D-2) K^{i_1}_bE^b_{i_1}K^{i_4}_{t_4}E^{t_4}_{i_4} - (D-1)K^{i_1}_bE^b_{i_1}K^{i_3}_dE^d_{i_3}) \\
  - \sqrt{|q|}R = \frac{-s}{\sqrt{q}} ( K^{i_1}_c K^{c}_{i_3}E_{d}^{i_1}E^d_{i_3} - K^{i_1}_bE^b_{i_1}K^{i_3}_dE^d_{i_3})  - \sqrt{|q|}R
\end{multline}
We rewrite the final result by using the same indices as in the book:
\begin{equation}
H =  \frac{-s}{\sqrt{|q|}} ( K^{l}_a K^{a}_{j}E_{l}^{b}E^j_{b} - K^{l}_aK^{j}_bE^a_{l}E^b_{j})  - \sqrt{|q|}R
\end{equation}

\section{Appendix C Timelike $SO(2,1)$ Symplectic Structure Calculations}
\label{sec:AppendixSymplecticStructure}

\textbf{Momentum-Momentum Poisson Bracket}

\begin{equation}
P^{ab}(x) = -|\det(E^c_e)|^{-2/(D-1)}( E^a_kE^{t}_{m}\eta^{mk}K^i_tE^b_i  - E^a_kE^{b}_{m}\eta^{mk}K^i_tE^t_i ) = \frac{1}{q}( E^a_kE^{t}_{m}\eta^{mk}K^i_tE^b_i  - E^a_kE^{b}_{m}\eta^{mk}K^i_tE^t_i )
\end{equation}

\begin{equation}
P^{cd}(x) = -|\det(E^c_e)|^{-2/(D-1)} ( E^c_kE^{t}_{m}\eta^{mk}K^i_tE^d_i  - E^c_kE^{d}_{m}\eta^{mk}K^i_tE^t_i)= \frac{1}{q}( E^c_kE^{t}_{m}\eta^{mk}K^i_tE^d_i  - E^c_kE^{d}_{m}\eta^{mk}K^i_tE^t_i)
\end{equation}

\begin{multline}
\label{momentumbracket2}
\{P^{ab}(x), P^{cd}(y)\} = \{\frac{1}{q}(E^a_{k_1}E^{t_1}_{p_1}\eta^{p_1k_1}K^{i_1}_{t_1}E^b_{i_1} - E^a_{k_2}E^{b}_{p_2}\eta^{p_2k_2}K^{i_2}_{t_2}E^{t_2}_{i_2}) , \\
 \frac{1}{q}(E^c_{m_1}E^{e_1}_{p_3}\eta^{p_3m_1}K^{j_1}_{e_1}E^d_{j_1} - E^c_{m_2}E^{d}_{p_4}\eta^{p_4m_2}K^{j_2}_{e_2}E^{e_2}_{j_2})
\end{multline}
 By introducing the following notations:\\[2ex]
$a = 1/q$ \\
$b = E^a_{k_1}E^{t_1}_{p_1}\eta^{p_1k_1}K^{i_1}_{t_1}E^b_{i_1}$ \\
$c = E^a_{k_2}E^{b}_{p_2}\eta^{p_2k_2}K^{i_2}_{t_2}E^{t_2}_{i_2}$  \\
$d = 1/q$ \\
$e = E^c_{m_1}E^{e_1}_{p_3}\eta^{p_3m_1}K^{j_1}_{e_1}E^d_{j_1}$ \\
$f = E^c_{m_2}E^{d}_{p_4}\eta^{p_4m_2}K^{j_2}_{e_2}E^{e_2}_{j_2}$\\[2ex]
We can rewrite ($\ref{momentumbracket2}$) as:
\begin{equation}
\label{momentumbracket4}
\{P^{ab}(x), P^{cd}(y)\} =\{a(b-c), d(e-f)\}
\end{equation}
or by using the Leibniz rule for the Poisson brackets:
\begin{multline}
\label{momentumbracket3}
\{P^{ab}(x), P^{cd}(y)\} =
( a (\{b, d\} -\{c, d\}) (e-f)+ 
\\
d(\{a, e\} - \{a, f\})(b-c) + 
\\
ad\{b, e\} - ad\{c, e\} - ad\{b, f\} + ad\{c, f\}) + 
\\
\{a, d\}(e-f)(b-c))
\end{multline}
The last term is zero since $\{a, d\} = \{(\det E)^\frac{-2}{D-1}, \{(\det E)^\frac{-2}{D-1}\} = 0$, as $\{E^a_j(x), E^b_k(y)\} = 0$

Let's calculate separately $\{b, f\}$, $\{c, f\}$, $\{c, e\}$, $\{b, e\}$, $\{a, e\}$, $\{a, f\}$, $\{b, d\}$, $\{c, d\}$
\begin{multline}
\label{be0}
\{b, e\} = \{E^a_{k_1}E^{t_1}_{p_1}\eta^{p_1k_1}K^{i_1}_{t_1}E^b_{i_1}, \;  E^c_{m_1} E^{e_1}_{p_3}\eta^{p_3m_1}K^{j_1}_{e_1}E^d_{j_1}\}=
\\
E^a_{k_1}E^{t_1}_{p_1}\eta^{p_1k_1}\{K^{i_1}_{t_1},E^c_{m_1}\}E^b_{i_1}E^{e_1}_{p_3}\eta^{p_3m_1}K^{j_1}_{e_1}E^d_{j_1} + \\
E^a_{k_1}E^{t_1}_{p_1}\eta^{p_1k_1}E^c_{m_1}\{K^{i_1}_{t_1},E^{e_1}_{p_3}\eta^{p_3m_1}\}E^b_{i_1}K^{j_1}_{e_1}E^d_{j_1} + \\
E^a_{k_1}E^{t_1}_{p_1}\eta^{p_1k_1}E^c_{m_1}E^{e_1}_{p_3}\eta^{p_3m_1}K^{j_1}_{e_1}\{K^{i_1}_{t_1},E^d_{j_1}\}E^b_{i_1} +\\
E^c_{m_1}E^{e_1}_{p_3}\eta^{p_3m_1}\{E^a_{k_1},K^{j_1}_{e_1}\}E^d_{j_1}E^{t_1}_{p_1}\eta^{p_1k_1}K^{i_1}_{t_1},E^b_{i_1} + \\
E^c_{m_1}E^{e_1}_{p_3}\eta^{p_3m_1}E^a_{k_1}\{E^{t_1}_{p_1}\eta^{p_1k_1}K^{j_1}_{e_1}\}E^d_{j_1}K^{i_1}_{t_1},E^b_{i_1} +\\
E^c_{m_1}E^{e_1}_{p_3}\eta^{p_3m_1}E^a_{k_1}E^{t_1}_{p_1}\eta^{p_1k_1}K^{i_1}_{t_1}\{E^b_{i_1}, K^{j_1}_{e_1}\}E^d_{j_1}
\end{multline}
\begin{multline}
\label{be}
\{b, e\} =  \{E^a_{k_1}E^{t_1}_{p_1}\eta^{p_1k_1}K^{i_1}_{t_1}E^b_{i_1}, \;  E^c_{m_1} E^{e_1}_{p_3}\eta^{p_3m_1}K^{j_1}_{e_1}E^d_{j_1}\}=
\\
E^a_{k_1}E^{t_1}_{p_1}\eta^{p_1k_1}(-\frac{k}{2}\delta^{i_1}_{m_1}\delta^c_{t_1})E^b_{i_1}E^{e_1}_{p_3}\eta^{p_3m_1}K^{j_1}_{e_1}E^d_{j_1} + \\
E^a_{k_1}E^{t_1}_{p_1}\eta^{p_1k_1}E^c_{m_1}(-\frac{k}{2}\delta^{i_1}_{m_1}\delta^{e_1}_{p_3}\eta^{p_3t_1})E^b_{i_1}K^{j_1}_{e_1}E^d_{j_1} + \\
E^a_{k_1}E^{t_1}_{p_1}\eta^{p_1k_1}E^c_{m_1}E^{e_1}_{p_3}\eta^{p_3m_1}K^{j_1}_{e_1}(-\frac{k}{2}\delta^{i_1}_{j_1}\delta^d_{t_1})E^b_{i_1} +\\
E^c_{m_1}E^{e_1}_{p_3}\eta^{p_3m_1}(\frac{k}{2}\delta^{j_1}_{k_1}\delta^a_{e_1})E^d_{j_1}E^{t_1}_{p_1}\eta^{p_1k_1}K^{i_1}_{t_1}E^b_{i_1} + \\
E^c_{m_1}E^{e_1}_{p_3}\eta^{p_3m_1}E^a_{k_1}(\frac{k}{2}\delta^{t_1}_{e_1}\delta^{j_1}_{p_1})\eta^{p_1k_1}E^d_{j_1}K^{i_1}_{t_1}E^b_{i_1} +\\
E^c_{m_1}E^{e_1}_{p_3}\eta^{p_3m_1}E^a_{k_1}E^{t_1}_{p_1}\eta^{p_1k_1}K^{i_1}_{t_1}(\frac{k}{2}\delta^{j_1}_{i_1}\delta^b_{e_1})E^d_{j_1} = 
\\
q^2\frac{k}{2}(-q^{ac}q^{be_1}K^{j_1}_{e_1}E^d_{j_1} - q^{ae_1}q^{bc}K^{j_1}_{e_1}E^d_{j_1} - q^{ad}q^{ce_1}K^{j_1}_{e_1}E^b_{j_1} + q^{ca}q^{dt_1}K^{i_1}_{t_1}E^b_{i_1} 
\\
+q^{ct_1}q^{da}K^{i_1}_{t_1}E^b_{i_1} + q^{cb}q^{at_1}K^{i_1}_{t_1}E^d_{i_1}) =
\\
=q^2\frac{k}{2}q( -q^{ac} \hat{G}^{bd} - q^{bc} \hat{G}^{ad} - q^{ad} \hat{G}^{cb} + q^{ac} \hat{G}^{db} + q^{da} \hat{G}^{cb} + q^{cb} \hat{G}^{ad}) = 
\\ = \frac{k}{2}q^3q^{ac}( \hat{G}^{db} - \hat{G}^{bd} ) = \frac{k}{2}q^3q^{ac}G^{db}
\end{multline}
,where
\begin{equation}
G^{db} := \hat{G}^{db} - \hat{G}^{bd} 
\end{equation}
and we have introduced the notations for $\hat{G}$ with various indices:
\begin{equation}
 \hat{G}^{db} = q^{de_1}K^{j_1}_{e_1}q^{bp}E^{j_1}_{p}  
\end{equation}
 Thus
\begin{equation}
\label{bfResult}
\{b, e\} =  \frac{k}{2}q^3q^{ac}G^{db} 
\end{equation}
The next Poisson bracket is:
\begin{multline}
\label{ae2}
\{a, e\} = \{(\det E)^\frac{-2}{D-1}, \;   E^c_{m_1}E^{e_1}_{p_3}\eta^{p_3m_1}K^{j_1}_{e_1}E^d_{j_1}\}=
\\
E^c_{m_1}E^{e_1}_{p_3}\eta^{p_3m_1}\{ (\det E)^\frac{-2}{D-1}, K^{j_1}_{e_1}\}E^d_{j_1} = 
\\
E^c_{m_1}E^{e_1}_{p_3}\eta^{p_3m_1}\frac{-2}{D-1}  (\det E)^\frac{-2}{D-1} \frac{\{ (\det E), K^{j_1}_{e_1}\}}{\det E}E^d_{j_1}=
\\
E^c_{m_1}E^{e_1}_{p_3}\eta^{p_3m_1}\frac{-2}{D-1} \frac{1}{q} E^n_r \{E^r_n, K^{j_1}_{e_1} \}E^d_{j_1} = 
\\
q^{ce_1} q \frac{-2}{D-1} \frac{1}{q}E^n_r (\frac{k}{2}\delta^{j_1}_n \delta_{e_1}^r) E^d_{j_1}=
\\
q^{ce_1} E^d_{j_1}\frac{-k}{D-1} E^{j_1}_{e_1} = q^{ce_1} \delta^d_{e_1} \frac{-k}{D-1} =  q^{cd}\frac{-k}{D-1}
\end{multline}
We obtain:
\begin{equation}
\label{aeResult1}
\{a, e\} =  q^{cd}\frac{-k}{D-1}
\end{equation}
The next bracket can be obtained from ($\ref{ae2}$), by changing the sign and making the following index replacement:
\begin{equation}
c \rightarrow a, \; d \rightarrow b, \; a \rightarrow c, \; b \rightarrow d
\end{equation}
\begin{equation}
\label{bd2}
\{b, d\} = \{E^a_{k_1}E^{t_1}_{p_1}\eta^{p_1k_1}K^{i_1}_{t_1}E^b_{i_1}, \; (\det E)^\frac{-2}{D-1}\}
\end{equation}
\begin{equation}
\label{bdResult2}
\{b, d\} =  q^{ab}\frac{k}{D-1}
\end{equation}
The next bracket goes as follows:
\begin{multline}
\{b, f\} = \{ E^a_{k_1}E^{t_1}_{p_1}\eta^{p_1k_1}K^{i_1}_{t_1}E^b_{i_1}, \; E^c_{m_2}E^{d}_{p_4}\eta^{p_4m_2}K^{j_2}_{e_2}E^{e_2}_{j_2}\}=
\\
E^a_{k_1}E^{t_1}_{p_1}\eta^{p_1k_1}\{K^{i_1}_{t_1},E^{c}_{m_2}\}E^b_{i_1}E^{d}_{p_4}\eta^{p_4m_2}K^{j_2}_{e_2}E^{e_2}_{j_2} + 
\\
E^a_{k_1}E^{t_1}_{p_1}\eta^{p_1k_1}E^c_{m_2}\{K^{i_1}_{t_1},E^{d}_{p_4}\eta^{p_4m_2}\}E^b_{i_1}K^{j_2}_{e_2}E^{e_2}_{j_2} + 
\\
E^a_{k_1}E^{t_1}_{p_1}\eta^{p_1k_1}E^c_{m_2}E^{d}_{p_4}\eta^{p_4m_2}K^{j_2}_{e_2}\{K^{i_1}_{t_1},E^{e_2}_{j_2}\}E^b_{i_1} + 
\\
E^c_{m_2}E^{d}_{p_4}\eta^{p_4m_2}\{E^{a}_{k_1}, K^{j_2}_{e_2}\}E^{e_2}_{j_2}E^{t_1}_{p_1}\eta^{p_1k_1}K^{i_1}_{t_1}E^{b}_{i_1} +
\\
E^c_{m_2}E^{d}_{p_4}\eta^{p_4m_2}E^a_{k_1}\{E^{t_1}_{p_1}\eta^{p_1k_1}, K^{j_2}_{e_2}\}E^{e_2}_{j_2}K^{i_1}_{t_1}E^{b}_{i_1} +
\\
E^c_{m_2}E^{d}_{p_4}\eta^{p_4m_2}E^a_{k_1}E^{t_1}_{p_1}\eta^{p_1k_1}K^{i_1}_{t_1}\{E^{b}_{i_1}, K^{j_2}_{e_2}\}E^{e_2}_{j_2}
\end{multline}
or
\begin{multline}
\label{bf4}
\{b, f\} = \{ E^a_{k_1}E^{t_1}_{p_1}\eta^{p_1k_1}K^{i_1}_{t_1}E^b_{i_1}, \; E^c_{m_2}E^{d}_{p_4}\eta^{p_4m_2}K^{j_2}_{e_2}E^{e_2}_{j_2}\} =
\\
E^a_{k_1}E^{t_1}_{p_1}\eta^{p_1k_1}(-\frac{k}{2}\delta^{i_1}_{m_2} \delta^{c}_{t_1} )E^b_{i_1}E^{d}_{p_4}\eta^{p_4m_2}K^{j_2}_{e_2}E^{e_2}_{j_2} + 
\\
E^a_{k_1}E^{t_1}_{p_1}\eta^{p_1k_1}E^c_{m_2}(-\frac{k}{2}\delta^{i_1}_{p_4} \delta^{d}_{t_1} )\eta^{p_4m_2}E^b_{i_1}K^{j_2}_{e_2}E^{e_2}_{j_2} + 
\\
E^a_{k_1}E^{t_1}_{p_1}\eta^{p_1k_1}E^c_{m_2}E^{d}_{p_4}\eta^{p_4m_2}K^{j_2}_{e_2}(-\frac{k}{2}\delta ^{i_1}_{j_2} \delta^{e_2}_{t_1})E^b_{i_1} + 
\\
E^c_{m_2}E^{d}_{p_4}\eta^{p_4m_2}(\frac{k}{2}\delta^{a}_{e_2} \delta^{j_2}_{k_1})E^{e_2}_{j_2}E^{t_1}_{p_1}\eta^{p_1k_1}K^{i_1}_{t_1}E^{b}_{i_1} +
\\
E^c_{m_2}E^{d}_{p_4}\eta^{p_4m_2}E^a_{k_1}(\frac{k}{2}\delta^{t_1}_{e_2} \delta^{j_2}_{p_1} )\eta^{p_1k_1}E^{e_2}_{j_2}K^{i_1}_{t_1}E^{b}_{i_1} +
\\
E^c_{m_2}E^{d}_{p_4}\eta^{p_4m_2}E^a_{k_1}E^{t_1}_{p_1}\eta^{p_1k_1}K^{i_1}_{t_1}(\frac{k}{2}\delta^{b}_{e_2} \delta^{j_2}_{i_1}\}E^{e_2}_{j_2} =
\\
\frac{k}{2}(- q^{ac}qq^{bd}qK^{j_2}_{e_2}E^{e_2}_{j_2} - q q^{ad}qq^{bc}K^{j_2}_{e_2}E^{e_2}_{j_2} - q^{ae_2}qq^{cd}qK^{j_2}_{e_2}E^{b}_{j_2} +
\\
q^{cd}qq^{at_1}qK^{i_1}_{t_1}E^b_{i_1} + q^{cd}qq^{at_1}qK^{i_1}_{t_1}E^b_{i_1} + 
q^{cd}qq^{at_1}qK^{i_1}_{t_1}E^b_{i_1} ) = 
\\
\frac{-k q^2}{2}K^{j_2}_{e_2}E^{e_2}_{j_2}(q^{ac}q^{bd} + q^{ad}q^{bc}) + kq^2q^{cd}q^{at_1}K^{i_1}_{t_1}E^b_{i_1}
\end{multline}
To summarize 
\begin{multline}
\label{bf4Result}
\{b, f\} =\{ E^a_{k_1}E^{t_1}_{p_1}\eta^{p_1k_1}K^{i_1}_{t_1}E^b_{i_1}, \; E^c_{m_2}E^{d}_{p_4}\eta^{p_4m_2}K^{j_2}_{e_2}E^{e_2}_{j_2}\}
\\
=-\frac{kq^2}{2}K^{j_2}_{e_2}E^{e_2}_{j_2}(q^{ac}q^{bd} + q^{ad}q^{bc}) + kq^2q^{cd}q^{at_1}K^{i_1}_{t_1}E^b_{i_1}
\\
= -\frac{kq^2}{2}K^{j_2}_{e_2}E^{e_2}_{j_2}(q^{ac}q^{bd} + q^{ad}q^{bc}) + kq^2q^{cd}q^{at_1}K^{i_1}_{t_1}qq^{bp}E^{i_1}_p
\\
=  -\frac{kq^2}{2}K^{j_2}_{e_2}E^{e_2}_{j_2}(q^{ac}q^{bd} + q^{ad}q^{bc}) + kq^3q^{cd}K^{ai_1}E^{bi_1}
\end{multline}
The bracket $\{c, e\}$ is similar to $\{f, b\} = -\{b, f\}$ above in ($\ref{bf4}$) with the following index replacement:
\begin{equation}
a \rightarrow c, \; b \rightarrow d, \; c \rightarrow a, \; d \rightarrow b
\end{equation}
\begin{multline}
\label{ce}
\{c, e\} = \{E^a_{k_2}E^{b}_{p_2}\eta^{p_2k_2}K^{i_2}_{t_2}E^{t_2}_{i_2},  \; E^c_{m_1}E^{e_1}_{p_3}\eta^{p_3m_1}K^{j_1}_{e_1}E^d_{j_1}\} = -\{E^c_{m_1}E^{e_1}_{p_3}\eta^{p_3m_1}K^{j_1}_{e_1}E^d_{j_1}, \; E^a_{k_2}E^{b}_{p_2}\eta^{p_2k_2}K^{i_2}_{t_2}E^{t_2}_{i_2}\}
\\
=\frac{kq^2}{2}K^{j_2}_{e_2}E^{e_2}_{j_2}(q^{ca}q^{db} + q^{da}q^{cb}) - kq^2q^{ab}q^{ct_1}K^{i_1}_{t_1}E^d_{i_1}
\\
=\frac{kq^2}{2}K^{j_2}_{e_2}E^{e_2}_{j_2}(q^{ca}q^{db} + q^{cb}q^{da}) - kq^3q^{ab}K^{ci_1}E^{di_1}
\end{multline}

\begin{multline}
\label{af6}
\{a, f\} = \{ {(\det E)}^{\frac{-2}{D-1}}, \; E^c_{m_2}E^{d}_{p_4}\eta^{p_4m_2}K^{j_2}_{e_2}E^{e_2}_{j_2}\} = 
E^c_{m_2}E^{d}_{p_4}\eta^{p_4m_2}\frac{-2}{D-1} \frac{1}{q} \frac{\{ (\det E), K^{j_2}_{e_2}\} }{(\det E)} E^{e_2}_{j_2}= 
\\
E^c_{m_2}E^{d}_{p_4}\eta^{p_4m_2}\frac{-2}{D-1} \frac{1}{q} E^{j_3}_{e_3}\{E^{e_3}_{j_3}, K^{j_2}_{e_2}\}E^{e_2}_{j_2} = 
E^c_{m_2}E^{d}_{p_4}\eta^{p_4m_2}\frac{-2}{D-1} \frac{1}{q} E^{j_3}_{e_3}\frac{k}{2}\delta^{j_2}_{j_3} \delta^{e_2}_{e_3}E^{e_2}_{j_2} =
\\
\frac{-k}{D-1} q q^{cd} \frac{1}{q} E^{j_2}_{e_2}E^{e_2}_{j_2} = \frac{-k}{D-1} q^{cd}D
\end{multline}
The bracket $\{c, d\}$ can be calculated from ($\ref{af6}$) by changing the sign and making the following index replacement:
\begin{equation}
c \rightarrow a, \; d \rightarrow b, \; a \rightarrow c, \; b \rightarrow d
\end{equation}
We  obtain:
\begin{equation}
\label{cd6}
\{c, d\} = \{E^a_{k_2}E^{b}_{p_2}\eta^{p_2k_2}K^{i_2}_{t_2}E^{t_2}_{i_2}, \; (\det E)^{\frac{-2}{D-1}}\} = \frac{k}{D-1}  q^{ab}D
\end{equation}
Finally we need to calculate the last bracket $\{c, f\}$:
\begin{multline}
\{c, f\} = \{E^a_{k_2}E^{b}_{p_2}\eta^{p_2k_2}K^{i_2}_{t_2}E^{t_2}_{i_2}, \; E^c_{m_2}E^{d}_{p_4}\eta^{p_4m_2}K^{j_2}_{e_2}E^{e_2}_{j_2}\}=
\\
E^a_{k_2}E^{b}_{p_2}\eta^{p_2k_2}\{K^{i_2}_{t_2}, E^c_{m_2}\}E^{t_2}_{i_2}E^{d}_{p_4}\eta^{p_4m_2}K^{j_2}_{e_2}E^{e_2}_{j_2} + 
\\
E^a_{k_2}E^{b}_{p_2}\eta^{p_2k_2}E^c_{m_2}\{K^{i_2}_{t_2}, E^{d}_{p_4}\eta^{p_4m_2}\}E^{t_2}_{i_2}K^{j_2}_{e_2}E^{e_2}_{j_2} + 
\\
E^a_{k_2}E^{b}_{p_2}\eta^{p_2k_2}E^c_{m_2}E^{d}_{p_4}\eta^{p_4m_2}K^{j_2}_{e_2}\{K^{i_2}_{t_2}, E^{e_2}_{j_2}\}E^{t_2}_{i_2} +
\\
E^c_{m_2}E^{d}_{p_4}\eta^{p_4m_2}\{E^a_{k_2}, K^{j_2}_{e_2}\}E^{e_2}_{j_2}E^{b}_{p_2}\eta^{p_2k_2}K^{i_2}_{t_2}E^{t_2}_{i_2} + 
\\
E^c_{m_2}E^{d}_{p_4}\eta^{p_4m_2}E^a_{k_2}\{E^{b}_{p_2}\eta^{p_2k_2}, K^{j_2}_{e_2}\}E^{e_2}_{j_2}K^{i_2}_{t_2}E^{t_2}_{i_2} + 
\\
E^c_{m_2}E^{d}_{p_4}\eta^{p_4m_2}E^a_{k_2}E^{b}_{p_2}\eta^{p_2k_2}K^{i_2}_{t_2}\{E^{t_2}_{i_2}, K^{j_2}_{e_2}\}E^{e_2}_{j_2}
\end{multline}
or
\begin{multline}
\{c, f\} = \{E^a_{k_2}E^{b}_{p_2}\eta^{p_2k_2}K^{i_2}_{t_2}E^{t_2}_{i_2}, \; E^c_{m_2}E^{d}_{p_4}\eta^{p_4m_2}K^{j_2}_{e_2}E^{e_2}_{j_2}\} =
\\
E^a_{k_2}E^{b}_{p_2}\eta^{p_2k_2}E^{t_2}_{i_2}(-\frac{k}{2}\delta^{i_2}_{m_2}\delta^c_{t_2})E^{d}_{p_4}\eta^{p_4m_2}K^{j_2}_{e_2}E^{e_2}_{j_2} + 
\\
E^a_{k_2}E^{b}_{p_2}\eta^{p_2k_2}E^{t_2}_{i_2}E^c_{m_2}(- \frac{k}{2}\delta^{i_2}_{p_4} \delta^d_{t_2})\eta^{p_4m_2}K^{j_2}_{e_2}E^{e_2}_{j_2} + 
\\
E^a_{k_2}E^{b}_{p_2}\eta^{p_2k_2}E^{t_2}_{i_2}E^c_{m_2}E^{d}_{p_4}\eta^{p_4m_2}K^{j_2}_{e_2}(- \frac{k}{2} \delta^{i_2}_{j_2}\delta^{e_2}_{t_2}) +
\\
E^c_{m_2}E^{d}_{p_4}\eta^{p_4m_2}( \frac{k}{2} \delta^{j_2}_{k_2}\delta^a_{e_2})E^{e_2}_{j_2}E^{b}_{p_2}\eta^{p_2k_2}K^{i_2}_{t_2}E^{t_2}_{i_2} + 
\\
E^c_{m_2}E^{d}_{p_4}\eta^{p_4m_2}E^a_{k_2}( \frac{k}{2} \delta^{j_2}_{p_2}\delta^b_{e_2})E^{e_2}_{j_2}\eta^{p_2k_2}K^{i_2}_{t_2}E^{t_2}_{i_2} + 
\\
E^c_{m_2}E^{d}_{p_4}\eta^{p_4m_2}E^a_{k_2}E^{b}_{p_2}\eta^{p_2k_2}K^{i_2}_{t_2}(\frac{k}{2} \delta^{t_2}_{e_2}\delta^{j_2}_{i_2})E^{e_2}_{j_2} = 
\\
\frac{k}{2}(-q^{ab}qq^{cd}qK^{j_2}_{e_2}E^{e_2}_{j_2} - q^{ab}qq^{cd}qK^{j_2}_{e_2}E^{e_2}_{j_2} - q^{ab}qq^{cd}qK^{j_2}_{e_2}E^{e_2}_{j_2} 
\\
+ q^{cd}qq^{ab}qK^{i_2}_{t_2}E^{t_2}_{i_2} +  q^{cd}qq^{ab}qK^{i_2}_{t_2}E^{t_2}_{i_2} +  q^{cd}qq^{ab}qK^{i_2}_{t_2}E^{t_2}_{i_2}) = 0
\end{multline}
Thus
\begin{equation}
\label{cf}
\{c, f\} = \{E^a_{k_2}E^{b}_{p_2}\eta^{p_2k_2}K^{i_2}_{t_2}E^{t_2}_{i_2}, \; E^c_{m_2}E^{d}_{p_4}\eta^{p_4m_2}K^{j_2}_{e_2}E^{e_2}_{j_2}\}=0
\end{equation}
To summarize we have obtained:\\[2ex]
$\{b, d\} =  \frac{k}{D-1}q^{ab}$\\
$\{c, d\} = \frac{k}{D-1} q^{ab}D$\\
$\{a, e\} =  \frac{-k}{D-1}q^{cd}$\\
$\{a, f\} = \frac{-k}{D-1} q^{cd}D$\\
$\{b, e\} = \frac{k}{2}qq^{ac}G^{db}$\\
$\{c, e\} = \frac{kq^2}{2}K^{j_2}_{e_2}E^{e_2}_{j_2}(q^{ca}q^{db} + q^{da}q^{cb}) - kq^3q^{ab}K^{ci_1}E^{di_1}$\\
$\{b, f\} = -\frac{kq^2}{2}K^{j_2}_{e_2}E^{e_2}_{j_2}(q^{ac}q^{bd} + q^{ad}q^{bc}) + kq^3q^{cd}K^{ai_1}E^{bi_1}$\\
$\{c, f\} = 0$\\[2ex]

By substituting into ($\ref{momentumbracket3}$) :
\begin{multline}
\label{momentumbracket7}
\{P^{ab}(x), P^{cd}(y)\} = ( a (\{b, d\} - \{c, d\})(e-f) + 
\\
d(\{a, e\} -\{a, f\})(b-c) + 
\\
ad\{b, e\} - ad\{c, e\} - ad\{b, f\} + ad\{c, f\})
\end{multline}
we obtain:
\begin{multline}
 \{P^{ab}(x), P^{cd}(y)\} = \frac{1}{q}(\frac{k}{D-1}q^{ab} - \frac{k}{D-1} q^{ab}D)(E^c_{m_1}E^{e_1}_{p_3}\eta^{p_3m_1}K^{j_1}_{e_1}E^d_{j_1} - E^c_{m_2}E^{d}_{p_4}\eta^{p_4m_2}K^{j_2}_{e_2}E^{e_2}_{j_2})
  + 
\\
\frac{1}{q} 
 (\frac{-k}{D-1}q^{cd} -  \frac{-k}{D-1} q^{cd}D) (E^a_{k_1}E^{t_1}_{p_1}\eta^{p_1k_1}K^{i_1}_{t_1}E^b_{i_1} - E^a_{k_2}E^{b}_{p_2}\eta^{p_2k_2}K^{i_2}_{t_2}E^{t_2}_{i_2}) 
 \\
 +\frac{k}{2}\frac{1}{q^2}  q^3q^{ac}G^{db} 
 \\-\frac{k}{2q^2} (q^2K^{j_2}_{e_2}E^{e_2}_{j_2}(q^{ca}q^{db} + q^{da}q^{cb}) - 2q^2q^{ab}q^{ct_1}K^{i_1}_{t_1}E^d_{i_1}) 
 \\
 -\frac{k}{2q^2} (-q^2K^{j_2}_{e_2}E^{e_2}_{j_2}(q^{ac}q^{bd} + q^{ad}q^{bc}) + 2q^2q^{cd}q^{at_1}K^{i_1}_{t_1}E^b_{i_1}))
 \end{multline}
\begin{multline}
 \{P^{ab}(x), P^{cd}(y)\} =  \frac{1}{q}(qq^{ce_1}K^{j_1}_{e_1}E^d_{j_1} - q q^{cd}K^{j_2}_{e_2}E^{e_2}_{j_2})(-k q^{ab})
+  \frac{1}{q}(qq^{at_1}K^{i_1}_{t_1}E^b_{i_1} - qq^{ab}K^{i_2}_{t_2}E^{t_2}_{i_2}))(k q^{cd} ) 
\\
+2kqq^{ac}G^{db} 
\\
+
\frac{k}{q^2}(q^2q^{ab}K^{ci_1}E^{d}_{i_1} -  q^2q^{cd}K^{ai_1}E^{b}_{i_1} )
\end{multline}
\begin{multline}
 \{P^{ab}(x), P^{cd}(y)\} =  -k q^{ab}K^{cj_1}E^d_{j_1}
+  k q^{cd} K^{ai_1}E^b_{i_1} 
+ 2kq q^{ac}G^{db} 
\\
+
k(q^{ab}K^{ci_1}E^{d}_{i_1} -  q^{cd}K^{ai_1}E^{b}_{i_1} ) = 2kqq^{ac}G^{bd}
\end{multline}

\begin{equation}
 \{P^{ab}(x), P^{cd}(y)\} = 2kqq^{ac}G^{bd}
\end{equation}\\[2ex]

\textbf{Coordinate-Momentum Poisson Bracket }\\[2ex]

We mark each line by the label L(line number) and provide the detailed comments underneath the formula on how we move from one line to the next in our calculations.
\begin{equation}
P^{ab}(x) = -|\det(E^c_e)|^{-2/(D-1)}( E^a_{k_1}E^{t_1}_{m_1}\eta^{m_1k_1}K^{i_1}_{t_1}E^b_{i_1} - E^a_{k_2}E^{b}_{m_2}\eta^{m_2k_2}K^{i_2}_{t_2}E^{t_2}_{i_2} ) 
\end{equation}
 \begin{equation}
 q_{cd}(y) = E^{i_3}_cE^{j_3}_d \eta_{i_3j_3}(-|\det(E^a_e)|^{2/(D-1)})
 \end{equation}
 \begin{multline}
 \label{FirstPoissonBook}
 \mbox{L1:  }\, \{P^{ab}(x),  q_{cd}(y)\} = |\det(E^c_e)|^{-2/(D-1)} \{( E^a_{k_1}E^{t_1}_{m_1}\eta^{m_1k_1}K^{i_1}_{t_1}E^b_{i_1} - E^a_{k_2}E^{b}_{m_2}\eta^{m_2k_2}K^{i_2}_{t_2}E^{t_2}_{i_2} ) , \; 
 \\ E^{i_3}_cE^{j_3}_d \eta_{i_3j_3}(\det(E))^{2/(D-1)}\} 
 \\
    \mbox{L2:  }\,=\frac{1}{q} \{ (E^a_{k_1}E^t_{m_1} \eta^{m_1k_1}  K^{i_1}_{t_1} qq^{be}E^{j_1}_e\eta_{i_1j_1}  - E^a_{k_2}E^b_{m_2} \eta^{m_2k_2} K^{i_2}_{t_2} qq^{t_2e}E^{j_2}_e\eta_{i_2j_2}),  \; E^{i_3}_cE^{j_3}_d\eta_{i_3j_3} (\det(E))^{2/(D-1)}\}
  \\
 =\frac{1}{q}q^2(q^{at}q^{be} - q^{ab}q^{te})E^j_e\eta_{ij}\{K^i_t, \; E^{i_3}_cE^{j_3}_d \eta_{i_3j_3} (\det(E))^{2/(D-1)}\}
\\
   \mbox{L3:  }\, =q(q^{at}q^{be} - q^{ab}q^{te})E^j_e\eta_{ij}(q\{K^i_tE^{i_3}_c\}E^{j_3}_d\eta_{i_3j_3} + q\{K^i_tE^{j_3}_d\}E^{i_3}_c\eta_{i_3j_3} + \frac{2}{D-1} \frac{q_{cd}}{q}q \frac{\{K^i_t, \det(E)\}}{\det(E)}) =
 \\
  \mbox{L4:  }\, q(q^{at}q^{be} - q^{ab}q^{te})E^j_e\eta_{ij}(q(-E^j_c)\{K^i_t, E^p_j\}E^{i_3}_pE^{j_3}_d\eta_{i_3j_3} + q(-E^j_d)\{K^i_t, E^p_j\}E^{j_3}_pE^{i_3}_c\eta_{i_3j_3}+
 \\
 \frac{2}{D-1} q_{cd} \{K^i_t, E^m_j\}E^j_m) = 
 \\
   \mbox{L5:  }\, q(q^{at}q^{be} - q^{ab}q^{te})E^j_e\eta_{ij}(q(-E^j_c)(-\frac{k}{2} \delta^i_j \delta^p_t) E^{i_3}_pE^{j_3}_d\eta_{i_3j_3} + q(-E^j_d)(-\frac{k}{2} \delta^i_j \delta^p_t) E^{j_3}_pE^{i_3}_c\eta_{i_3j_3}+
 \\
 \frac{2}{D-1} q_{cd} (-\frac{k}{2} \delta^i_j \delta^m_t ) E^j_m) = 
 \\
  \mbox{L6:  }\, kq(q^{at}q^{be} - q^{ab}q^{te})E^j_e\eta_{ij}(qE^i_cE^{i_3}_tE^{j_3}_d\eta_{i_3j_3} + qE^i_dE^{j_3}_tE^{i_3}_c\eta_{i_3j_3} - \frac{2}{D-1} q_{cd} E^i_t) =
  \\
   \mbox{L7:  }\,  kq(q^{at}q^{be} - q^{ab}q^{te})(q \frac{q_{ec}}{q} \frac{ q_{dt}}{q} + q \frac{q_{ed}}{q} \frac{q_{ct}}{q} - 
   \frac{2}{D-1} q_{cd} \frac{q_{et}}{q}) =
   \\
   \mbox{L8:  }\,  k (q^{at}q^{be} - q^{ab}q^{te})(q_{ec}q_{dt} + q_{ed}q_{ct} - \frac{2}{D-1} q_{cd}q_{et})=
   \\
    \mbox{L9:  }\, k(\delta^b_c \delta^a_d + \delta^b_d \delta^a_c) - 2q^{ab}q_{cd}q - \frac{2}{D-1}(q_{cd}q^{ab} - D q_{cd}q^{ab})=
   \\
 \mbox{L10:  }\, k(\delta^b_c \delta^a_d + \delta^b_d \delta^a_c) - 2q^{ab}q_{cd}q - \frac{2}{D-1}(1 - D )q_{cd}q^{ab} = 
   \\
  \mbox{L11:  }\, k(\delta^b_c \delta^a_d + \delta^b_d \delta^a_c) - 2q^{ab}q_{cd}q + 2q^{ab}q_{cd}q = 
   \\
   \mbox{L12:  }\,  k\delta^b_{(c}\delta^a_{d)}\delta(x, y)
 \end{multline}
 \begin{equation}
  \{P^{ab}(x),  q_{cd}(y)\} =   k\delta^b_{(c}\delta^a_{d)}\delta(x, y)
 \end{equation}
 ,where\\
 in the line $\mbox{L2:}$ we used  $E^a_i E^b_j \eta^{ij} = q q^{ab}$ and $ q := (\det(E))^{2/(D-1)}$\\
 in the line  $\mbox{L3:}$ we used Leibniz rule and $\; \mbox{and} \{E^a_j(x), E^b_k(y)\} = 0$\\
 in the line $\mbox{L4:}$ we used: $\delta E^i_a = -E^i_a \delta E^b_i E^i_b$ and $ [\delta(E)]/\det(E) = E^j_a\delta E^a_j$\\
 in the line $\mbox{L5:}$ we calculated the Poisson brackets: $ \{E^a_i(x), K^j_b(y)\} = \frac{k}{2}\delta^a_b\delta^j_i\delta(x, y)$\\
 in the line $\mbox{L7:}$ we used: $E^i_e E^j_c \eta_{ij}= q_{ec}/q $, etc, \\
 in the line $\mbox{L9}$: we have opened the parentheses and used:
 $q^{at} q_{td} = \delta^a_d$ and  $q^{et} q_{et} = D$\\
in the  line $\mbox{L10:}$ $D-1$ cancels.\\
in the  line $\mbox{L11:}$ the last two terms are the same and mutually cancel. 

\section{Appendix D Timelike $SO(2,1)$ Riemann Curvature Expression in $SO(2,1)$ case}
\label{sec:RiemannCurvatureExpression}
In this Appendix we derive the following two formulas:
\begin{equation}
\label{Riemann2}
R^i_{ab} = 2\partial_{[a}\Gamma^i_{b]} + \bar{\eta}^{ij}{\epsilon_{jkl}}_{so(3)}\Gamma^{k}_a\Gamma^l_b
\end{equation}
and
\begin{equation}
\label{Riemann3}
^{(\beta)}F^i_{ab} = 2\partial_{[a} ^{(\beta)}A^i_{b]} +  \bar{\eta}^{ij}{\epsilon_{jkl}}_{so(3)} \; ^{(\beta)}A^k_a  \; ^{(\beta)}  A^l_b
\end{equation}
 , where $\bar{\eta}_{ij} = Diag(1,1,1)$  in $SO(3)$ case, and $\bar{\eta}_{ij} = Diag(-1,1,1)$ in $SO(2,1)$ case.
 We begin with the curvature tensor definition:
 \begin{equation}
 \label{RiemannCurvature}
 R^a_{bcd} = \partial_b \Gamma^a_{cd}  - \partial_c \Gamma^a_{bd} + \Gamma^a_{bt}\Gamma^t_{cd} - \Gamma^a_{ct} \Gamma^t_{bd}
 \end{equation}
\textbf{$SO(3)$ case:}\\[2ex]
In $SO(3)$ case ($\ref{Riemann2}$) becomes:
\begin{equation}
\label{Riemann11}
R^i_{ab} = 2\partial_{[a}\Gamma^i_{b]} + {\epsilon^i_{kl}}_{so(3)}\Gamma^{k}_a\Gamma^l_b
\end{equation}
We then use $so(3) \rightarrow R^3$ isomorphism:
\begin{equation}
\label{isomophism4}
 \Gamma^l_{ai}  = {\epsilon_{ki}^{l}}_{so(3)} \Gamma^k_a
\end{equation}
By using the torsion-free condition: $\Gamma^l_{ai} = \Gamma^l_{ia}$, we can write it also as:
\begin{equation}
\label{isomophism5}
 \Gamma^l_{ai}  =  \Gamma^l_{ia} =  {\epsilon_{ki}^{l}}_{so(3)} \Gamma^k_a = {\epsilon_{ka}^{l}}_{so(3)} \Gamma^k_i
\end{equation}
By substituting ($\ref{isomophism5}$) into ($\ref{RiemannCurvature}$), we obtain:
\begin{equation}
\label{Riemann5}
R^a_{bcd} = \partial_b ({\epsilon^a_{pd}}_{so(3)}\Gamma^p_c) -\partial_c ({\epsilon^a_{pd}}_{so(3)} \Gamma^p_b) +
{\epsilon^a_{pt}}_{so(3)}\Gamma^p_b{\epsilon^t_{sd}}_{so(3)}\Gamma^s_c - {\epsilon^a_{mt}}_{so(3)}\Gamma^m_c {\epsilon^t_{ld}}_{so(3)}\Gamma^l_b
\end{equation}
We rewrite the third term by using the antisymmetric tensor properties:
\begin{equation}
\label{ThirdTerm}
{\epsilon^a_{pt}}_{so(3)}{\epsilon^t_{sd}}_{so(3)}  = {\epsilon^{tap}}_{so(3)}{\epsilon_{tsd}}_{so(3)}  = 2\delta^{[a}_s\delta^{p]}_d = 
\delta^a_s\delta^p_d - \delta^p_s\delta^a_d
\end{equation}
and the fourth term by:
\begin{equation}
\label{FourthTerm}
{\epsilon^a_{mt}}_{so(3)}{\epsilon^t_{ld}}_{so(3)} = {\epsilon^{tam}}_{so(3)}{\epsilon_{tld}}_{so(3)} = 2\delta^{[a}_l\delta^{m]}_d = 
\delta^a_l\delta^m_d - \delta^m_l\delta^a_d
\end{equation}
we obtain:
\begin{multline}
\label{Riemann6}
R^a_{bcd} = {\epsilon^a_{pd}}_{so(3)}( \partial_b \Gamma^p_c -\partial_c \Gamma^p_b  )+
(\delta^a_s\delta^p_d - \delta^p_s\delta^a_d)\Gamma^p_b\Gamma^s_c -(\delta^a_l\delta^m_d - \delta^m_l\delta^a_d) \Gamma^m_c \Gamma^l_b =
\\
{\epsilon^a_{pd}}_{so(3)}( \partial_b \Gamma^p_c -\partial_c \Gamma^p_b  )+ ( \Gamma^a_c\Gamma^d_b - \Gamma^s_b\Gamma^s_c) - (\Gamma^d_c\Gamma^a_b  - \Gamma^l_b\Gamma^l_c) =
\\
{\epsilon^a_{pd}}_{so(3)}( \partial_b \Gamma^p_c -\partial_c \Gamma^p_b  ) +  \Gamma^a_c\Gamma^d_b -  \Gamma^a_b\Gamma^d_c 
\end{multline}
By renaming index $p \rightarrow i$:
\begin{equation}
\label{Riemann7}
R^a_{bcd} = 
{\epsilon^a_{id}}_{so(3)}( \partial_b \Gamma^i_c -\partial_c \Gamma^i_b  ) + \Gamma^d_b \Gamma^a_c -\Gamma^a_{b}\Gamma^d_c  
\end{equation}
In order to prove formula ($\ref{Riemann11}$) we will show that if we take $R^i_{bc}$ in that form and contract it with ${\epsilon^a_{id}}_{so(3)}$ we will obtain $R^a_{bcd}$ as in ($\ref{Riemann7}$):
\begin{multline}
\label{Riemann10}
{\epsilon^a_{id}}_{so(3)} R^i_{bc} = {\epsilon^a_{id}}_{so(3)}( \partial_b \Gamma^i_c -\partial_c \Gamma^i_b  ) + {\epsilon^a_{id}}_{so(3)} {\epsilon^i_{kl}}_{so(3)}\Gamma^{k}_b\Gamma^l_c = 
\\
{\epsilon^a_{id}}_{so(3)}( \partial_b \Gamma^i_c -\partial_c \Gamma^i_b  ) - {\epsilon^a_{di}}_{so(3)} {\epsilon^i_{kl}}_{so(3)}\Gamma^{k}_b\Gamma^l_c = 
\\
 {\epsilon^a_{id}}_{so(3)}( \partial_b \Gamma^i_c -\partial_c \Gamma^i_b  ) - 2\delta^{[a}_k\delta^{d]}_l\Gamma^{k}_b\Gamma^l_c =  
 \\
 {\epsilon^a_{id}}_{so(3)}( \partial_b \Gamma^i_c -\partial_c \Gamma^i_b  ) -  \Gamma^{a}_b\Gamma^d_c + \Gamma^{d}_b\Gamma^a_c  
\end{multline}
By comparing it to ($\ref{Riemann7}$) we see that it equals to $R^a_{bcd}$. So, we have proved that:
\begin{equation}
R^a_{bcd} =  {\epsilon^a_{id}}_{so(3)} R^i_{bc}
\end{equation}
, where $R^i_{ab} = 2\partial_{[a}\Gamma^i_{b]} + {\epsilon^i_{kl}}_{so(3)}\Gamma^{k}_a\Gamma^l_b$\\[2ex]
Let us now consider $SO(2,1)$ case:\\[2ex]
We use the $so(2,1) \rightarrow R^3_{2,1}$ isomorphism:
\begin{equation}
\label{isomophism6}
 \Gamma^l_{\;ai}  = {\epsilon_{\;ki}^{l}}_{so(2,1)} \Gamma^k_a
\end{equation}
So we are getting $SO(2,1)$ case from $SO(3)$ case very easily by replacing everywhere ${\epsilon_{\;ki}^{l}}_{so(3)}$ tensors with ${\epsilon_{\;ki}^{l}}_{so(2,1)}$. The only difference that we will encounter is a different sign compared to ($\ref{ThirdTerm}$) and ($\ref{FourthTerm}$):
\begin{equation}
\label{ThirdTerm1}
{\epsilon^a_{pt}}_{so(2,1)}{\epsilon^t_{sd}}_{so(2,1)}  = {\epsilon^{tap}}_{so(2,1)}{\epsilon_{tsd}}_{so(2,1)}  = -2\delta^{[a}_s\delta^{p]}_d = 
-(\delta^a_s\delta^p_d - \delta^p_s\delta^a_d)
\end{equation}
and the fourth term by:
\begin{equation}
\label{FourthTerm1}
{\epsilon^a_{mt}}_{so(2,1)}{\epsilon^t_{ld}}_{so(2,1)} = {\epsilon^{tam}}_{so(2,1)}{\epsilon_{tld}}_{so(2,1)} = - 2\delta^{[a}_l\delta^{m]}_d = 
- (\delta^a_l\delta^m_d - \delta^m_l\delta^a_d)
\end{equation}
It only changes the sign of the last two terms in the $so(2,1)$ analog of ($\ref{Riemann7}$) and of ($\ref{Riemann10}$)
\begin{equation}
\label{Riemann12}
R^a_{\;bcd} = 
{\epsilon^a_{\;id}}_{so(2,1)}( \partial_b \Gamma^i_c -\partial_c \Gamma^i_b  ) + \Gamma^a_{b}\Gamma^d_c - \Gamma^d_b \Gamma^a_c
\end{equation}
\begin{equation}
\label{Riemann13}
{\epsilon^a_{\;id}}_{so(2,1)} R^i_{bc} =
 {\epsilon^a_{\;id}}_{so(2,1)}( \partial_b \Gamma^i_c -\partial_c \Gamma^i_b  ) + \Gamma^{a}_b\Gamma^d_c  -  \Gamma^{d}_b\Gamma^a_c 
\end{equation}
So again we obtain:
\begin{equation}
\label{Riemann14}
R^a_{\;bcd} =  {\epsilon^a_{\;id}}_{so(2,1)} R^i_{\;bc}
\end{equation}
, where $R^i_{ab} = 2\partial_{[a}\Gamma^i_{b]} + {\epsilon^i_{\;kl}}_{so(2,1)}\Gamma^{k}_a\Gamma^l_b$\\[2ex]
By combining two cases and remembering that we can express: $ {\epsilon^i_{\;kl}}_{so(2,1)} = \bar{\eta}^{ij} {\epsilon_{jkl}}_{so(3)}$ we can write it in a general case:
\begin{equation}
R^i_{ab} = 2\partial_{[a}\Gamma^i_{b]} + {\bar{\eta}}^{ij}{\epsilon_{jkl}}_{so(3)} \Gamma^{k}_a\Gamma^l_b
\end{equation}
, which finally proves formula ($\ref{Riemann2}$).
In addition the general form of ($\ref{Riemann14}$) is as follows:
\begin{equation}
\label{333}
R^a_{\;bcd} =  {\bar{\eta}}^{at}{\epsilon_{tid}}_{so(3)} R^i_{\;bc} 
\end{equation}
If we use generalized Sen-Ashtekar connection from ($\ref{GeneralizedConnection}$):
\begin{equation}
\label{GeneralizedConnection1}
^{\beta}A^k_a = \Gamma^k_a + (^{\beta}K^k_a),   \;\;\; ^{\beta}A^l_{ai} = \bar{\eta}_{ij}{\epsilon_k^{jl}}_{so(3)} A^k_a
\end{equation}
and repeat all the steps using $A^k_a$ instead of $\Gamma^k_a$, we will obtain ($\ref{Riemann3}$).

\section{Appendix E  Expressing   $^{(\beta)}F^j_{ab} $ via $R^j_{ab} $}
\label{sec:ExpressingFviaR}

\begin{equation}
\label{Riemann15}
R^i_{ab} = 2\partial_{[a}\Gamma^i_{b]} + \bar{\eta}^{ij}{\epsilon_{jkl}}_{so(3)}\Gamma^{k}_a\Gamma^l_b
\end{equation}
We want to prove that:
\begin{equation}
\label{Riemann16}
^{(\beta)}F^i_{ab} = 2\partial_{[a} ^{(\beta)}A^i_{b]} +  \bar{\eta}^{ij}{\epsilon_{jkl}}_{so(3)} \; ^{(\beta)}A^k_a  \; ^{(\beta)}  A^l_b
\end{equation}
Proof:\\[2ex]
 We remind that $^{(\beta)} E^b_i= E^b_i / \beta $ and $^{(\beta)} K^k_a= \beta K^k_a $ \\[2ex]
By substituting  $^{(\beta)}A^k_a =\Gamma^k_a + \beta K^k_{a}$ into ($\ref{Riemann16}$) we obtain:
\begin{multline}
\label{Riemann17}
^{(\beta)}F^i_{ab} = 2\partial_{[a}\Gamma^i_{b]} + 2\partial_{[a} ^{(\beta)}K^i_{b]} + \bar{\eta}^{ij}{\epsilon_{jkl}}_{so(3)} \; (\Gamma^k_a + \beta K^k_{a})(\Gamma^l_b + \beta K^l_{b}) = 
\\
(2\partial_{[a}\Gamma^i_{b]} + \bar{\eta}^{ij}{\epsilon_{jkl}}_{so(3)} \; \Gamma^k_a  \Gamma^l_b) + (2\partial_{[a} ^{(\beta)}K^i_{b]} +  \bar{\eta}^{ij}{\epsilon_{jkl}}_{so(3)}\Gamma^k_a  \beta K^l_{b} + \bar{\eta}^{ij}{\epsilon_{jkl}}_{so(3)} \Gamma^l_b  \beta K^k_{a})\\
 +  (\beta^2 \bar{\eta}^{ij}{\epsilon_{jkl}}_{so(3)} K^k_a K^l_b) =  R^i_{ab} + 2\beta D_{[a} K^i_{b]} + \beta^2 \bar{\eta}^{ij}{\epsilon_{jkl}}_{so(3)} K^k_a K^l_b
\end{multline}
Therefore:
\begin{equation}
\label{ExtrinsicCurvature2}
^{(\beta)}F^i_{ab} = R^i_{ab} + 2\beta D_{[a} K^i_{b]} + \beta^2 \bar{\eta}^{ij}{\epsilon_{jkl}}_{so(3)} K^k_a K^l_b
\end{equation}

\section{Appendix F Timelike $SO(2,1)$ Diffeomorphism Contraction Calculations}
\label{sec:DiffeomorphismContractionCalculatons}

We would like to derive the following expression:
\begin{equation}
\label{Contracting1}
^{(\beta)}F^i_{ab} \;  ^{(\beta)} E^b_i = \frac{R^i_{ab} E^b_i}{\beta}  + 2D_{[a}K^i_{b]}E^b_i + \beta K^i_aG_i
\end{equation}
We begin with  ($\ref{ExtrinsicCurvature2}$):
\begin{equation}
\label{ExtrinsicCurvature101}
^{(\beta)}F^i_{ab} = R^i_{ab} + 2\beta D_{[a} K^i_{b]} + \beta^2 \bar{\eta}^{ij}{\epsilon_{jkl}}_{so(3)} K^k_a K^l_b
\end{equation} 
contracting it with  $^{(\beta)}E^b_j$ 
\begin{equation}
\label{Contracting2}
^{(\beta)}F^i_{ab} \;  ^{(\beta)} E^b_i = \frac{R^i_{ab} E^b_i}{\beta}  + 2\beta D_{[a}K^i_{b]} \frac{E^b_i}{\beta} + \beta^2 \bar{\eta}^{ij}{\epsilon_{jkl}}_{so(3)} K^k_a K^l_b \frac{E^b_i}{\beta}
\end{equation}
by simplifying it we obtain:
\begin{equation}
\label{Contracting2}
^{(\beta)}F^i_{ab} \;  ^{(\beta)} E^b_i = \frac{R^i_{ab} E^b_i}{\beta}  + 2 D_{[a}K^i_{b]} E^b_i + \beta \bar{\eta}^{ij}{\epsilon_{jkl}}_{so(3)} K^k_a K^l_b E^b_i
\end{equation}
So, we only need to prove that the last term is a rotational constraint, i.e:
\begin{equation}
\bar{\eta}^{ij}{\epsilon_{jkl}}_{so(3)} K^k_a K^l_b E^b_i = K^i_aG_i
\end{equation}
We do it in the following steps:
\begin{multline}
\bar{\eta}^{ij}{\epsilon_{jkl}}_{so(3)} K^k_a K^l_b E^b_i =  K^k_a ( {\epsilon_{jkl}}_{so(3)}K^l_b E^{bj} ) = K^k_a ( {\epsilon_k^{lj}}_{so(3)}K_{lb} E^b_j )  \\
=  K^k_a ( \bar{\eta}_{km} {\epsilon^{mlj}}_{so(3)}K_{lb} E^b_j ) = K^k_a G_k = K^i_aG_i
\end{multline}
,where in the first step above we used generalized metric to raise index $j$, in the next step we raised and lowered at the same time indices $lj$, and permutated twice $jkl \rightarrow klj$, so the sign stays the same, then we raised index m by using generalized metric, since in $so(3)$ case it is all the same, while in $so(2,1)$ case even  ${\epsilon_{jkl}}_{so(3)}$ indices should be lowered and raised by using Minkowski tensor $\eta_{ij}$. Finally we used the following rotational constraint expression obtained in ($\ref{RotConstraint2}$):
\begin{equation}
\label{RotConstraint3}
G_i = \bar{\eta}_{ij}{\epsilon^{jkl}}_{so(3)}K_{ak}E^a_{l}
\end{equation}

\section{Appendix G Identity from Rotational Constraint}
\label{sec:GIdentityfromRotational}
\begin{equation}
\label{rconstraint3}
G_{ab} := K^i_{[a}E^j_{b]}\bar{\eta}_{ij} = 1/2( K_{ai}E^i_b -  K_{bi}E^i_a )= 0
\end{equation}
therefore:
\begin{equation}
\label{rconstraint4}
  K_{bi}E^i_a  =  K_{ai}E^i_b 
\end{equation}
or, by raising index i
\begin{equation}
K^i_{b}E^i_{a} =  K^{i}_aE^i_{b} 
\end{equation}
it follows that 
\begin{equation}
 K^i_{b}E^a_{i} =  K^{i}_aE^b_{i} 
\end{equation}
since 
\begin{equation}
\label{146}
K^i_aE^b_i = K^i_aE^i_tq^{tb}q
\end{equation}
,while
\begin{equation}
\label{147}
K^i_bE^a_i = K^i_bE^i_tq^{ta}q
\end{equation}
and the right hand sides of ($\ref{146}$) and ($\ref{147}$) are equal, as
\begin{equation}
K^i_aE^i_tq^{tb}= K^i_bE^i_tq^{ta}
\end{equation}
since
\begin{equation}
K^i_a = K^i_b q^{ta} q_{tb}= K^i_b \delta^{a}_b
\end{equation}

\section{Appendix I Contracting Riemann Curvature with Triads Second Term}
\label{sec:HContractingWithTriadsSecondTerm}
We prove the following identity first in $SO(3)$ spacelike foliation ADM case:
\begin{equation}
2\beta D_{[a} K^j_{b]} {\epsilon_j^{\;\;kl}}_{so(3)} {^{(\beta)}E^{a}_{k}}{^{(\beta)}E^{b}_{l}} = - 2 ^{(\beta)}E^a_kD_aG^k
\end{equation}
it can be rewritten as:
\begin{multline}
\label{1010}
2\beta D_{[a} K^j_{b]} {\epsilon_j^{\;\;kl}}_{so(3)} {^{(\beta)}E^{a}_{k}}{^{(\beta)}E^{b}_{l}} = \\
\beta( D_aK^j_b - D_bK^j_a) {\epsilon_j^{\;\;kl}}_{so(3)} {^{(\beta)}E^{a}_{k}}{^{(\beta)}E^{b}_{l}} = \\
{^{(\beta)}E^{a}_{k}}  ( D_a (K^j_b E^{b}_{l}) - {^{(\beta)}E^{b}_{l}} D_b(K^j_a E^{a}_{k})){\epsilon_j^{\;\;kl}}_{so(3)}  = \\
-{^{(\beta)}E^{a}_{k}}  ( D_a (K^j_b E^{b}_{l}){\epsilon_j^{kl}}_{so(3)} - {^{(\beta)}E^{b}_{l}} D_b(K^j_a E^{a}_{k})){\epsilon_j^{lk}}_{so(3)}  =\\
 -{^{(\beta)}E^{a}_{k}} D_a(G^k) -  {^{(\beta)}E^{b}_{l}} D_b(G^l) =
\\
-2{^{(\beta)}E^{a}_{k}} D_a(G^k) 
\end{multline}
,where in the fourth line for the first and second terms we used rotational constraint definition ($\ref{RotConstraint8}$): $G^k = {\epsilon_j^{kl}}_{so(3)}K^j_bE^b_{l}$ and moved index $k$  by one position for the first term: ${\epsilon_j^{\;\;kl}}_{so(3)} = -{\epsilon_j^{kl}}_{so(3)}$ and index $l$ by two positions for the second term ${\epsilon_j^{\;\;kl}}_{so(3)} = {\epsilon_j^{lk}}_{so(3)}$
It is easy to pass to generalized form:
\begin{equation}
2\beta D_{[a} K^j_{b]}  \bar{\eta}_{ji}{\epsilon^{ikl}}_{so(3)} {^{(\beta)}E^{a}_{k}}{^{(\beta)}E^{b}_{l}} = - 2 ^{(\beta)}E^a_kD_aG^k
\end{equation}
Similar to the above:
\begin{multline}
\label{1010}
2\beta D_{[a} K^j_{b]}  \bar{\eta}_{ji}{\epsilon^{ikl}}_{so(3)} {^{(\beta)}E^{a}_{k}}{^{(\beta)}E^{b}_{l}} = \\
\beta( D_aK^j_b - D_bK^j_a)  \bar{\eta}_{ji}{\epsilon^{ikl}}_{so(3)} {^{(\beta)}E^{a}_{k}}{^{(\beta)}E^{b}_{l}} = \\
{^{(\beta)}E^{a}_{k}}  ( D_a (K^j_b E^{b}_{l}) - {^{(\beta)}E^{b}_{l}} D_b(K^j_a E^{a}_{k})) \bar{\eta}_{ji}{\epsilon^{ikl}}_{so(3)}  = \\
-{^{(\beta)}E^{a}_{k}}  ( D_a (K^j_b E^{b}_{l})\bar{\eta}_{ji}{\epsilon^{kil}}_{so(3)} - {^{(\beta)}E^{b}_{l}} D_b(K^j_a E^{a}_{k}))\bar{\eta}_{ji}{\epsilon^{lik}}_{so(3)}  =\\
 -{^{(\beta)}E^{a}_{k}} D_a(G^k) -  {^{(\beta)}E^{b}_{l}} D_b(G^l) =
\\
-2{^{(\beta)}E^{a}_{k}} D_a(G^k) 
\end{multline}
, where we used generalized rotational constraint ($\ref{RotConstraint8}$):
\begin{equation}
\label{RotConstraint80}
G_i = \bar{\eta}_{ij}{\epsilon^{jkl}}_{so(3)}K_{ak}E^a_{l}=\bar{\eta}_{ij}{\epsilon_k^{jl}}_{so(3)}K^k_{a}E^a_{l}=0
\end{equation}
and
\begin{equation}
\label{RotConstraint81}
G_i = \bar{\eta}_{ij}{\epsilon^{jkl}}_{so(3)}K_{ak}E^a_{l}=\bar{\eta}_{ij}{\epsilon_k^{jl}}_{so(3)}K^k_{a}E^a_{l}=0
\end{equation}
since $G_i=0$, $G^k= \bar{\eta}^{ki}G_i=0$
\begin{equation}
G^k= \bar{\eta}^{ki}G_i= K^j_b E^{b}_{l}\bar{\eta}_{ji}{\epsilon^{kil}}_{so(3)} = 0
\end{equation}

\section{Appendix J Contracting Riemann Curvature with Triads Third Term}
\label{sec:HContractingWithTriadsThirdTerm}
We would like to prove that in $SO(3)$ spacelike ADM case:
\begin{equation}
  {\beta^2 }{\epsilon^j_{\;mn}}_{so(3)} K^m_a K^n_b {\epsilon_j^{\;\;kl}}_{so(3)} {^{(\beta)}E^{a}_{k}}{^{(\beta)}E^{b}_{l}} = (K^j_aE^a_j)^2 - (K^j_bE^a_j)(K^k_aE^b_k) 
\end{equation}
The proof is straightforward. $\beta$ cancels on the left hand side right away and we use ${\epsilon^j_{\;mn}}_{so(3)}{\epsilon_j^{\;\;kl}}_{so(3)} = 2 \delta^{[k}_m \delta^{l]}_n$:
\begin{multline}
  {\beta^2 }{\epsilon^j_{\;mn}}_{so(3)} K^m_a K^n_b {\epsilon_j^{\;\;kl}}_{so(3)} {^{(\beta)}E^{a}_{k}}{^{(\beta)}E^{b}_{l}} = K^m_a K^n_b  2\delta^{[k}_m \delta^{l]}_n E^{a}_{k}E^{b}_{l} =\\
 (\delta^k_m\delta^l_n - \delta^k_n \delta^l_m)  K^m_a K^n_b  E^{a}_{k}E^{b}_{l} =   ( K^k_a K^l_b  E^{a}_{k}E^{b}_{l} - K^l_a K^k_b  E^{a}_{k}E^{b}_{l}) = \\
( (K^k_aE^a_k)^2 - (K^l_bE^a_l))(K^k_aE^b_k) = ((K^j_aE^a_j)^2 - (K^j_bE^a_j)(K^k_aE^b_k))
\end{multline}
On the other hand, in $SO(2,1)$ case we have an opposite sign in:\\[2ex]
${\epsilon^j_{\;mn}}_{so(2,1)}{\epsilon_j^{\;\;kl}}_{so(2,1)} = -2 \delta^{[m}_k \delta^{n]}_l$:\\[2ex]
Therefore in $SO(2,1)$ timelike ADM case the sign also becomes opposite:
\begin{equation}
  {\beta^2 }{\epsilon^j_{\;mn}}_{so(2,1)} K^m_a K^n_b {\epsilon_j^{\;\;kl}}_{so(2,1)} {^{(\beta)}E^{a}_{k}}{^{(\beta)}E^{b}_{l}} = - ((K^j_aE^a_j)^2 - (K^j_bE^a_j)(K^k_aE^b_k))
\end{equation}
The generalized formula for both cases would look like this:
\begin{equation}
  {\beta^2 }\bar{\eta}^{ji}{\epsilon_{imn}}_{so(3)} K^m_a K^n_b \bar{\eta}_{jp}{\epsilon^{pkl}}_{so(2,1)} {^{(\beta)}E^{a}_{k}}{^{(\beta)}E^{b}_{l}} = -s ((K^j_aE^a_j)^2 - (K^j_bE^a_j)(K^k_aE^b_k))
\end{equation}
,where $s = -1$ in $SO(3)$ spacelike ADM case and $s=1$ in $SO(2,1)$ timelike foliation ADM case, $\bar{\eta}^{ij}$ - generalized metric : Diag(1,1,1) in spacelike and Diag(-1,1,1) in timelike foliation cases.

\section{Appendix K Contracting Riemann Curvature with one Triad}
\label{sec:HContractingRiemannWithOneTriad}
We need to prove that: 
\begin{equation}
\label{Contracting55}
R^i_{ab} E^b_i = 0
\end{equation}
The proof for $SO(3)$ spacelike foliation can be found in \cite{Thiemann} (4.2.35):
The Bianchi identity can be written in the form:
\begin{equation}
\label{Bianchi1}
{\epsilon_{ijk}}_{so(3)}{\epsilon^{efc}}_{so(3)}R^j_{ef}e^k_c=0 \rightarrow \frac{1}{2}{\epsilon_{ijk}}_{so(3}{\epsilon^{efc}}_{so(3)}R^j_{ef}e^k_ce^i_a = \frac{1}{2}E^b_j{\epsilon_{cab}}_{so(3)}{\epsilon^{efc}}_{so(3)}R^j_{ae} = R^j_{ab}E^b_j = 0
\end{equation}
It is still true in $SO(2,1)$ timelike ADM foliation case, as what changes are antisymmetric tensors ${\epsilon_{ijk}}_{so(3)} \rightarrow {\epsilon_{ijk}}_{so(2,1)}$, and it causes only a sign change in the last term when contracting $so(2,1)$ antisymmetric tensors: ${\epsilon_{cab}}_{so(2,1)}{\epsilon^{efc}}_{so(2,1)}= -2 \delta^{[e}_a\delta^{f]}_b$ instead of $so(3)$ version ${\epsilon_{cab}}_{so(3)}{\epsilon^{efc}}_{so(3)}= 2\delta^{[e}_a\delta^{f]}_b$
\begin{equation}
\label{Bianchi2}
{\epsilon_{ijk}}_{so(2,1)}{\epsilon^{efc}}_{so(2,1)}R^j_{ef}e^k_c=0 \rightarrow \frac{1}{2}{\epsilon_{ijk}}_{so(2,1}{\epsilon^{efc}}_{so(2,1)}R^j_{ef}e^k_ce^i_a = \frac{1}{2}E^b_j{\epsilon_{cab}}_{so(2,1)}{\epsilon^{efc}}_{so(2,1)}R^j_{ae} = -R^j_{ab}E^b_j = 0
\end{equation}

\end{document}